\documentclass{article}
\usepackage[T2A]{fontenc}
\usepackage[cp1251]{inputenc}
\usepackage[english,russian]{babel}
\usepackage[tbtags]{amsmath}
\usepackage{amsfonts,amssymb,mathrsfs,amscd}

\numberwithin{equation}{section}
\newtheorem{theorem}{Теорема}

\newtheorem{proof}{Доказательство}

\def\R{{\Bbb R}}
\def\C{{\Bbb C}}

\def\Tr{{\rm Tr}}

\def\T{{\rm T}}

\def\Mat{\mathrm{Mat}}

\def\diag{{\rm diag}}
\def\cl{{\cal C}\!\ell_{1,3}}
\def\Ccl{\C\otimes\cl}
\def\Even{{\rm Even}}

\def\tr{{\rm tr}}
\def\Tr{{\rm Tr}}

\def\M{{\mathcal M}}
\def\Id{{\rm Id}}

\newcommand{\be}{\begin{equation}}
\newcommand{\ee}{\end{equation}}
\def\st{\stackrel}
\def\SU{{\rm SU}}
\def\U{{\rm U}}
\def\O{{\rm O}}
\def\SO{{\rm SO}}
\def\su{{\rm su}}
\def\u{{\rm u}}

\def\Herm{{\rm Herm}}

\def\MC{{\rm Mat}(2,\C)}

\begin{document}

\title{Эскиз калибровочной модели гравитации с SU(2) симметрией на лоренцевом многообразии с тетрадой}
\author{Н.\,Г.~Марчук\footnote{Математический институт им.~В.\,А.~Стеклова РАН, Email: {nmarchuk@mi-ras.ru}
}}

\date{18.02.2024}

\maketitle

%\begin{fulltext}

\begin{abstract}
Предлагается калибровочная модель с SU(2) симметрией для описания гравитационного взаимодействия фундаментальных фермионов (лептонов и кварков) на  лоренцевом многообразии с тетрадой.  От системы уравнений Дирака-Янга-Миллса, лежащей в основе Стандартной Модели,  переходим к модельной системе уравнений Дирака-Ланцоша-Янга-Миллса, записываемой с помощью матриц второго порядка. Эта система уравнений обладает дополнительной калибровочной симметрией по отношению к унитарной группе  SU(2). Поле Янга-Миллса, ассоциированное с этой калибровочной группой,  интерпретируется как гравитационное поле, взаимодействующее с фундаментальными фермионами.

Библиография: 22 наименований.
\end{abstract}

%\begin{keywords}
Ключевые слова:
Общая Теория Относительности, гравитация, лоренцево многообразие, тетрада, матрицы Паули, уравнение Дирака, уравнения Янга-Миллса, уравнение Дирака-Ланцоша, калибровочная симметрия, примитивное полевое уравнение
%\end{keywords}

УДК {514.8}

\markright{Эскиз калибровочной модели гравитации}

%%%%%%%%%%%%%%%%%%%%%%%%%%%%%%%%%%%%%%%%%%%%%%%%%

\section*{Введение}

Задача объединения теории гравитации (основанной на Общей Теории Относительности (ОТО) А.~Эйнштейна или ее модификации) и Стандартной Модели элементарных частиц в единую теорию интересует исследователей уже долгое время. Теория квантовой гравитации должна рассматривать гравитационное взаимодействие на уровне элементарных частиц. В литературе отражено много подходов к созданию теории квантовой гравитации (см., например, обзоры \cite{Oriti,Rovelli}).  В настоящее время ни один из подходов к построению квантовой гравитации не признан научным сообществом полностью удовлетворительным -- теория квантовой гравитации еще не создана.

Пионерская работа 1956 года Р.~Утиямы \cite{Utiyama} (и его последователей) положила начало одному из направлений создания квантовой гравитации, а именно полевой теории гравитации с калибровочной симметрией. С современным состоянием работ указанного направления квантовой гравитации можно познакомиться, например, по  \cite{BlagoevicHehl}, \cite{KKrasnov}.

Предлагаемый в данной статье эскиз модели гравитации можно отнести к этому направлению полевых теорий гравитации с калибровочной симметрией, причем все калибровочные симметрии нашей модели являются унитарными (в отличие от модели Утиямы).
\bigskip

%%%%%%%%%%%%%%%%%%%%%%%%%%%%%%%%%%%%%%%%%%%%%%%%%%%%%%

 Рассматриваемая модель базируется на модельном уравнении Дирака, концепция которого разрабатывалась в серии работ автора, опубликованных в 2000--2009 годах (см., в частности, \cite{paper2002}) и подытоженных в монографии \cite{Marchuk2018}.

 Волновая функция фундаментальных фермионов в модельном уравнении Дирака описывается тензорными (а не спинорными) величинами,
обладающими дополнительной симметрией по отношению к
псевдоунитарной (конформной) группе $\SU(2,2)$, либо к одной из ее подгрупп
(группа $\SU(2,2)$ размерности 15 среди своих подгрупп
содержит симплектическую группу ${\rm Sp}(2,\R)$ размерности 10, спинорную группу ${\rm Spin}(1,3)$ размерности 6 и две унитарные подгруппы -- $\SU(2)\times\SU(2)$  размерности 6 и $\SU(2)$ размерности 3). Псевдоунитарная
симметрия является внутренней симметрией модельных уравнений и
никак не связана с заменами координат (пространства Минковского,
либо псевдориманова многообразия).

В связи с появлением нового уравнения (модельного уравнения Дирака) встал вопрос -- можно ли соотнести спинорные решения стандартного уравнения Дирака (1928) и тензорные решения модельного уравнения Дирака. Положительный ответ на этот вопрос был найден (см. \cite{paper2002,Marchuk2018}) и,  вкратце, заключается в следующем. Тензоры и спиноры являются
разными алгебро-геометрическими объектами, которые, вообще говоря,
не сводятся друг к другу. Это ясно видно в пространстве
Минковского, где тензоры и спиноры реализуют разные (тензорные и
спинорные) представления группы Лоренца.
%Данное замечание относится к общему понятию тензора и
%общему понятию спинора.
Но если рассматривать не просто тензоры, а тензорные решения
модельного уравнения Дирака с псевдоунитарной симметрией (либо с симплектической симметрией, либо со спинорной симметрией), то
ситуация изменится. Из любого тензорного решения модельного
уравнения Дирака можно получить спинорное решение с помощью
процедуры {\em спиноризации}. А именно, каждую лоренцеву
замену координат пространства Минковского надо сопроводить
соответствующим (синхронизованным) преобразованием из спинорной
подгруппы псевдоунитарной группы симметрии. То есть процедуру спиноризации можно рассматривать как своеобразный метод фиксации спинорной калибровочной симметрии модельного уравнения Дирака.

Таким образом приходим к модельному уравнению Дирака (а также к
модельным системам уравнений Дирака--Максвелла и
Дирака--Янга--Миллса), которые можно рассматривать как обобщения
соответствующих стандартных (спинорных) уравнений теории поля.

Следует отметить, что в работах упомянутого цикла 2000--2009 годов не удалось выделить одной группы симметрии (среди подгрупп группы Ли $\SU(2,2)$), которая была бы ``лучше'' других для рассматриваемой теории. В то время на первом плане был вопрос о связи между решениями стандартного (спинорного) уравнения Дирака и решениями модельного (тензорного) уравнения Дирака. Поэтому в качестве ``наиболее подходящих'' кандидатов на правильную группу симметрии были выделены три группы -- $\SU(2,2)$, ${\rm Sp}(2,\R)$, ${\rm Spin}(1,3)$ как группы, допускающие процедуру спиноризации
\footnote{
В тот период (2000--2009 годы) вопрос о неоднозначности группы симметрии мной ясно осознавался, но не вызывал большого беспокойства. Считал, что в процессе дальнейшего развития модели найдутся соображения, которые позволят однозначно выделить подходящую группу симметрии. В конечном счете так и произошло, но времени на решение вопроса ушло гораздо больше, чем я ожидал, и ответ оказался не таким, как я в то время ожидал. Правильной группой симметрии (как сейчас считаю) оказалась унитарная группа $\SU(2)$ и этот факт связан с новым направлением развития модели теории поля, которое было найдено в июне 2020 года.}.

Процедура спиноризации, это лишь один из возможных способов использования дополнительной калибровочной симметрии модельного уравнения Дирака. Другим возможным способом использования дополнительной калибровочной симметрии модельного уравнения Дирака является создание калибровочной модели гравитации с группой симметрии $\SU(2)$ (отметим, что группа калибровочной симметрии $\SU(2)$ уже встречалась в калибровочных моделях гравитации \cite{KKrasnov}).

Первоначальная обкатка этой идеи содержится в статье \cite{Marchuk_AACA2021}, где дан эскиз калибровочной модели гравитации с $\SU(2)$ симметрией в пространстве Минковского  $\R^{1,3}$ (сравнение подходов статьи \cite{Marchuk_AACA2021} и настоящей статьи обсуждается в параграфе \ref{par:AACA}).

Так как рассмотрение велось в плоском пространстве Минковского, то ни о какой реалистичности модели говорить не приходится. В указанной статье используется технический арсенал алгебры Клиффорда $\cl$ и ее комплексификации $\C\otimes\cl$. В частности, используются тензорные поля со значениями в $\C\otimes\cl$. Волновые функции лептонов описываются функциями от $x\in\R^{1,3}$ со значениями в левом идеале комплексифицированной алгебры Клиффорда. Гравитационное поле описывается как поле Янга-Миллса с унитарной $\SU(2)$ калибровочной симметрией. Причем в правой части уравнений Янга-Миллса стоит неабелев ток, который может рассматриваться (в некотором смысле) как квадратный корень из метрического тензора.

Настоящая статья содержит существенное развитие результатов статьи \cite{Marchuk_AACA2021} в следующих направлениях:
\begin{enumerate}
\item Модель переносится на лоренцево многообразие с выделенным времениподобным векторным полем и с гладким тетрадным полем.
\item От техники алгебры Клиффорда $\cl$ мы переходим к технике алгебры комплексных  матриц второго порядка $\Mat(2,\C)$. Соответственно, от модельного уравнения Дирака мы переходим к модельному уравнению Дирака-Ланцоша.
\item Рассматривая волновые функции лептонов, мы переходим от левого идеала алгебры $\C\otimes\cl$ (зависящего от восьми комплексных функций) к матрицам из $\Mat(2,\C)$ (зависящих от четырех комплексных функций).
\item Предложено обобщенное примитивное полевое уравнение (входящее в основную систему полевых уравнений) в правой части которого стоит выражение, зависящее от волновой функции фермиона.
\end{enumerate}

\medskip

 В параграфе 1 зафиксирована терминология и обозначения используемой в статье техники матриц из $\Mat(2,\C)$. В параграфе 2 рассматриваются лоренцевы многообразия с дополнительными структурами -- с выделенным времениподобным векторным полем и с согласованным гладким тетрадным полем. Рассматриваются элементы анализа  на лоренцевом многообразии, причем основными объектами рассмотрения являются тензорные поля со значениями в алгебре матриц  $\Mat(2,\C)$. Понятие ковариантной производной со связностью Леви-Чивиты обобщается на тензорные поля со значениями в $\Mat(2,\C)$. В параграфах 4-8 рассматриваются основные уравнения модельной теории поля на лоренцевом многообразии (с дополнительными структурами)  с $\SU(2)$, $\U(2)$ и $\SU(2)$, $\U(2)$, $\U(3)$ калибровочными симметриями соответственно. Каждая из вводимых систем уравнений, в частности, обладает локальной симметрией по отношению к группе специальных унитарных матриц второго порядка $\SU(2)$, которую мы интерпретируем как группу калибровочной симметрии гравитационного поля.
   В параграфе 6 рассматривается гамильтонов вид модельного уравнения Дирака на лоренцевом многообразии.
 В параграфе 9  проводим сравнение модельных уравнений теории поля рассматриваемых в настоящей статье с модельными уравнениями из статьи \cite{Marchuk_AACA2021}.
 \medskip

 В предлагаемой модели используется ряд результатов и идей, сформировавшихся усилиями многих исследователей в релятивистской квантовой физике (в частности, в теории калибровочных полей) и в теории гравитации.  Представляется, что модель содержит оригинальное сочетание известных и новых идей (таких как модельное уравнение Дирака-Ланцоша с двумя унитарными неабелевыми калибровочными симметриями, примитивное полевое уравнение и др.) и это дает надежду на то, что при дальнейшей разработке модель разовьется до полноценной теории поля.

Отметим, что все физические константы пока взяты равными единице.

%%%%%%%%%%%%%%%%%%%%%%%%%%%%%%%%%%%%

\section{Алгебра комплексных матриц,\\ эрмитовы и антиэрмитовы матрицы}
В этом параграфе зафиксируем стандартные обозначения и термины матричной алгебры, которые будут использоваться в рассматриваемой далее модели.

Пусть $n$ -- натуральное число и $\Mat(n,\C)$ -- алгебра квадратных $n\!\times\!n$-матриц с комплексными элементами.

Пусть $\Id_n$ -- единичная матрица порядка $n$.
Операция взятия следа матрицы  позволяет определить два оператора проектирования $\pi_\pm : \Mat(n,\C)\to\Mat(n,\C)$
\begin{eqnarray*}
\pi_+(M) &:=& (\frac{1}{n}\tr M) \Id_n,\\
\pi_-(M) &:=& M - \pi_+(M).
\end{eqnarray*}
При этом
$$
M = \pi_+(M) + \pi_-(M) = M_+ + M_-,\quad \forall M\in\Mat(n,\C),
$$
матрица $M_+=\pi_+(M)$  пропорциональна единичной матрице, а  $M_-=\pi_-(M)$ -- бесследовая матрица ($\tr M_- =0$).
Соответственно $n^2$-мерное комплексное векторное пространство матриц $\Mat(n,\C)$ можно представить в виде прямой суммы подпространств
$$
\Mat(n,\C)=\Mat_+(n,\C)\oplus\Mat_-(n,\C),
$$
где $\Mat_+(n,\C)$ -- одномерное комплексное векторное пространство матриц, пропорциональных единичной матрице, а $\Mat_-(n,\C)$ есть $(n^2-1)$-мерное комплексное векторное пространство бесследовых матриц.
Для матриц из $\Mat(n,\C)$ введем операцию сопряжения
\begin{equation}
M\to\tilde M=\pi_+(M)-\pi_-(M)=M_+-M_-,\quad \forall M\in\Mat(n,\C).\label{tilde0}
\end{equation}
Очевидно, что $\tilde{\tilde M}=M$ и
$$
\pi_+(M)=\frac{1}{2}(M+\tilde M),\quad \pi_-(M)=\frac{1}{2}(M-\tilde M).
$$

 Матрица $M\in\Mat(n,\C)$ называется {\em эрмитовой}, если $M^\dagger=M$, и {\em антиэрмитовой}, если $M^\dagger=-M$, где $\dagger$ --операция эрмитова сопряжения матрицы (матрица транспонируется и берется комплексное сопряжение от всех ее элементов). Если $M$ -- эрмитова матрица, то $i M$ -- антиэрмитова матрица. Любую матрицу можно представить в виде суммы эрмитовой и антиэрмитовой матриц
$$
V=\frac{1}{2}(V+V^\dagger) + \frac{1}{2}(V-V^\dagger),\quad \forall V\in\Mat(n,\C).
$$
Все собственные числа эрмитовой матрицы вещественные, а антиэрмитовой матрицы -- чисто мнимые. Любая эрмитова или антиэрмитова матрица приводится к диагональному виду  (соответственно с вещественными или чисто мнимыми элементами на диагонали) с помощью преобразования подобия
$$
M\to S^{-1}M S
$$
с соответствующей матрицей $S\in{\rm GL}(n,\C)$.

Эрмитовы матрицы можно рассматривать как  обобщение (на матрицы из $\Mat(n,\C)$) вещественных чисел, а антиэрмитовы  матрицы рассматриваются как обобщение чисто мнимых комплексных чисел. Операцию эрмитова сопряжения матрицы $\dagger$ можно рассматривать как обобщение операции сопряжения комплексного числа.

Множество эрмитовых матриц и множество антиэрмитовых матриц из $\Mat(n,\C)$ можно рассматривать как вещественные векторные пространства размерности $n^2$.
Множества антиэрмитовых и бесследовых антиэрмитовых матриц из $\Mat(n,\C)$ обозначаются
\begin{eqnarray}
\u(n) &=&\{M\in\Mat(n,\C) : M^\dagger=-M\},\label{un}\\
\su(n) &=&\{M\in u(n) : \tr\,M=0\}.\nonumber
\end{eqnarray}
Размерность вещественного векторного пространства $su(n)$ равна $n^2-1$.
Векторное пространство $\u(n)$ представляется в виде прямой суммы подпространств
$$
\u(n) = \u(1)\oplus \su(n),
$$
где\footnote{Согласно определению (\ref{un}), множество $\u(1)$ является множеством антиэрмитовых матриц размера $1$, т.е. состоит из чисто мнимых чисел $i\lambda$, $\lambda\in\R$. Для наших целей удобно вложить $\u(1)$ в алгебру матриц $\Mat(n,\C)$ сопоставляя $i\lambda\to i\lambda I\!d_n$.} $\u(1)\subset \u(n)$ рассматривается как одномерное подпространство элементов вида $i\lambda I\!d_n$, $\lambda\in\R$, $I\!d_n$ -- единичная матрица порядка $n$.
Множества $n\times n$-матриц $\u(n)$, $\su(n)$, $\u(1)$ замкнуты относительно коммутатора $[A,B]=AB-BA$ и могут рассматриваться как вещественные алгебры Ли соответствующих групп Ли унитарных $n\times n$-матриц
\begin{eqnarray*}
\U(n) &=&\{M\in\Mat(n,\C) : M^\dagger=M^{-1}\},\\
\SU(n) &=&\{M\in U(n) : \det\,M=1\},\\
\U(1) &=& \{e^{i\rho}I\!d_n : \rho\in\R\}.
\end{eqnarray*}

Рассмотрим еще некоторый группы Ли матриц третьего и четвертого порядков. Начнем с группы ортогональных и специальных ортогональных матриц третьего порядка
\begin{eqnarray*}
\O(3) &=& \{M\in\Mat(3,\R) : M^\T M=Id_3\},\\
\SO(3) &=& \{M\in O(3) : \det(M)=1\},
\end{eqnarray*}
где $M^\T$ -- транспонированная матрица.

Введем обозначение для  диагональной матрицы Минковского
$$
\eta=\|\eta_{ab}\|=\|\eta^{ab}\|={\rm diag}(1,-1,-1,-1)\in\Mat(4,\R),
$$
где элементы этой матрицы нумеруются индексами $a,b=0,1,2,3$.

Определим группы (по умножению) псевдоортогональных матриц четвертого порядка
\begin{eqnarray}
\O(1,3) &=& \{ M\in\Mat(4,\R) : M^\T\eta M=\eta\},\nonumber\\
\SO_+(1,3) &=& \{ M\in O(1,3) : \det(M)=1,\ m^0_0\geq1\},\label{SOplusStar}\\
\SO_+^*(1,3) &=& \{ M\in SO_+(1,3) : m^0_0=1,\ m^0_k=m^k_0=0,\ k=1,2,3\},\nonumber
\end{eqnarray}
где $m^a_b$ -- элементы матрицы $M$ и $a,b=0,1,2,3$.

Матрицы из группы $\SO_+^*(1,3)$ являются блочно-диагональными матрицами вида
$$
\begin{pmatrix}
1 & 0 & 0 & 0\cr
0 & p^1_1 & p^1_2 & p^1_3\cr
0 & p^2_1 & p^2_2 & p^2_3\cr
0 & p^3_1 & p^3_2 & p^3_3
\end{pmatrix}
$$
и матрица $P=\|p^k_l\|\in \SO(3)$. Группа $\SO_+^*(1,3)$ изоморфна группе $\SO(3)$.

%%%%%%%%%%%%%%%%%%%%%%%%%%%%%%%%%%%%%%%%%%%%%%%
\medskip

\noindent{\bf Алгебра матриц второго порядка $\Mat(2,\C)$}.
Дальше рассматриваем комплексные квадратные матрицы второго порядка. Единичную матрицу второго порядка будем обозначать
$$
e := Id_2.
$$
Для  матриц второго порядка введенная в (\ref{tilde0}) операция сопряжения имеет свойства
\begin{equation}
\widetilde{U V}=\tilde{V}\tilde{U},\quad \forall U,V\in\Mat(2,\C).\label{prop:1}
\end{equation}
\begin{equation}
V\tilde{V} = \tilde{V}V=\det(V) e,\quad \forall V\in\Mat(2,\C)\label{prop:2}
\end{equation}
и, если $\det(V)\neq0$, то
\begin{equation}
V^{-1} = \frac{1}{\det(V)}\tilde{V}.\label{prop:3}
\end{equation}
Отметим, что для матриц размера более двух соотношения (\ref{prop:1}), (\ref{prop:2}), (\ref{prop:3}), вообще говоря, не выполняются.

Введем обозначение для множества эрмитовых  матриц второго порядка
\begin{eqnarray*}
\Herm(2) &=& \{M\in\Mat(2,\C) : M^\dagger=M\},
%i\,\Herm(2) &=& \{M\in\Mat(2,\C) : M^\dagger=-M\}.
\end{eqnarray*}
Если $M_1,M_2\in\Herm(2)$, $\lambda_1,\lambda_2\in\R$, то $\lambda_1 M_1+\lambda_2 M_2\in\Herm(2)$. Поэтому множество эрмитовых матриц $\Herm(2)$ и множество антиэрмитовых матриц $i\,\Herm(2)$ являются вещественнымы векторнымы пространствами размерности $4$.
%Для дальнейшего, алгебру матриц $\Mat(2,\C)$ удобно рассматривать как комплексификацию $\C\times\X$ векторного пространства $\X$ (этот вопрос подробно обсуждается в дополнении, где вводятся операции йорданова и векторного умножения элементов $\X$).

Пусть $\sigma^0=e$ -- единичная матрица второго порядка и $\sigma^1,\sigma^2,\sigma^3$ -- матрицы Паули
\begin{equation}
\sigma^0 =\begin{pmatrix}1 & 0\\ 0 & 1 \end{pmatrix},\quad
\sigma^1 =\begin{pmatrix}0 & 1\\ 1 & 0 \end{pmatrix},\quad
\sigma^2 =\begin{pmatrix}0 &-i\\ i & 0 \end{pmatrix},\quad
\sigma^3 =\begin{pmatrix}1 & 0\\ 0 & -1 \end{pmatrix}.\label{Pauli1}
\end{equation}
Четыре эрмитовы матрицы (\ref{Pauli1}) называются {\em базисом Паули} пространства эрмитовых матриц $\Herm(2)$ и пространства комплексных матриц $\Mat(2,\C)$. Произвольную матрицу  $V\in\Mat(2,\C)$ можно записать в виде разложения по базису Паули
$$
V=v_k\sigma^k,\quad v_0,v_1,v_2,v_3\in\C.
$$
Для эрмитово сопряженной матрицы получаем формулу
$$
V^\dagger=\bar v_k\sigma^k,
$$
где $\bar v_0,\bar v_1,\bar v_2,\bar v_3$ -- комплексно сопряженные числа.

%%%%%%%%%%%%%%%%%%%%%%%%%%%%%%%%%%%%%%

В нижеследующих рассмотрениях латинские индексы $a,b,\ldots$ пробегающие значения $0,1,2,3$ можно поднимать или опускать с помощью матрицы Минковского $\eta=\|\eta_{ab}\|=\|\eta^{ab}\|$. Например $\sigma_a=\eta_{ab}\sigma^b$.

Для матриц из $\Mat(2,\C)$ определено эрмитово полуторалинейное скалярное произведение
\begin{equation}
(U,V) :=\tr(U^\dagger V)=\bar u_0v_0+\bar u_1v_1+\bar u_2v_2+\bar u_3v_3,\label{eu:scalar1:multz}
\end{equation}
где $u_a,v_a$ -- коэффициенты в разложении матриц $U,V$ по  базису Паули.

Эрмитова норма матрицы дается формулой
$$
\|U\|^2:=(U,U)=|u_0|^2+|u_1|^2+|u_2|^2+|u_3|^2,
$$
где $|u_a|$ -- модуль комплексного числа $u_a\in\C$.

Таким образом, введенное  скалярное произведение превращает алгебру матриц $\Mat(2,\C)$ в четырехмерное унитарное пространство.

Наконец, отметим, что алгебра матриц $\Mat(2,\C)$ изоморфна алгебре бикватернионов \cite{Kravchenko}.

%August 12, 2021

\medskip

\noindent{\bf Некоторые свойства матриц Паули.} Напомним, что для матриц второго порядка оператор $\pi_+=\frac{1}{2}\ \tr$.
 Легко проверить следующие свойства матриц $\sigma^a$:
\begin{eqnarray}
\frac{1}{2}\tr(\sigma^a\tilde\sigma_b) &=& \frac{1}{2}\tr(\tilde\sigma^a\sigma_b)=\delta^a_b,\label{sig:prop:1}\\
\frac{1}{2}\sigma^a A\sigma_a &=&\frac{1}{2}\tilde\sigma^a A\tilde\sigma_a=-\tilde A=-A_+ + A_- ,\label{sig:prop:2}\\
\frac{1}{2}\tilde\sigma^a A\sigma_a &=&\frac{1}{2}\sigma^a A\tilde\sigma_a=(\tr\,A)\sigma^0=2 A_+,\label{sig:prop:3}
\end{eqnarray}

%%%%%%%%%%%%%%%%%%%%%%%%%%%%%%%%%%%%%%%%%%%%%%%%%%%%%%%%

%%%%%%%%%%%%%%%%%%%%%%%%%%%%%%%%%%%%%%%%%%%%%

\section{Лоренцево многообразие}

Пусть $M$ есть четырехмерное дифференцируемое ориентируемое многообразие с локальными координатами $x^\mu$, $\mu=0,1,2,3$ и с метрическим тензором (тензорным полем) %$g = g_{\mu\nu} dx^\mu\otimes dx^\nu$,
$g_{\mu\nu}=g_{\nu\mu}$ сигнатуры $(1,3)$ (для любого $x\in M$ матрица $\|g_{\mu\nu}\|$ имеет одно положительное  и три отрицательных собственных числа). На многообразии $M$ допускаются общие гладкие замены координат с ненулевым якобианом. Комплект $(M,g)$ называется {\em лоренцевым многообразием}\footnote{Лоренцевы многообразия составляют подкласс псевдоримановых многообразий.}.

В ОТО А.~Эйнштейна физическое пространство-время моделируется лоренцевым многообразием, причем гравитационное поле моделируется метрическим тензором $g_{\mu\nu}$.
\medskip

\noindent{\bf Дополнительные условия на лоренцево многообразие.}
В предлагаемой модели физическое пространство-время моделируется лоренцевым многообразием $(M,g)$, удовлетворяющим следующим дополнительным условиям:

1) На многообразии допускаются только общие гладкие замены координат с положительным якобианом. Поэтому в модели псевдотензоры неотличимы от тензоров.

2) На лоренцевом многообразии $(M,g)$, в дополнение к метрическому тензорному полю $g_{\mu\nu}$, задано гладкое времениподобное нормированное векторное поле $\tau^\mu$, удовлетворяющее условиям\footnote{В \cite{Lich} показано, что на псевдоримановых многообразиях с сигнатурой $(+--\ldots-)$ всегда можно глобально задать  временниподобное векторное поле.
}
\begin{eqnarray}
%&& \tau^0 > 0,\label{tau:cond0}\\
&& \tau^\mu\tau_\mu=1,\label{tau:cond1}\\
&& \nabla_\mu\tau^\mu=0,\label{tau:cond2}
\end{eqnarray}
где $\tau_\mu=g_{\mu\nu}\tau^\nu$ и $\nabla_\mu$ -- ковариантная производная (связность Леви-Чивиты).
\medskip

%!!!!!!!!!!!!!!! Векторное поле $\tau^\mu$ фактически требуется для того, чтобы ввести класс $\tau$-стабилизированных тетрад. Больше он вроде нигде не используется. Нет не так - используется

В моделях гравитации лоренцевы многообразия с заданным гладким времениподобным векторным полем рассматривались многими авторами в основном в связи с вопросами задания систем отсчета в ОТО (см. обзор в \cite{VladUS1982}).

%где $\tau=\tau_\mu dx^\mu$ и $\delta=*d*$ -- оператор обобщенной дивергенции\footnote{В случае пространства Минковского условие (\ref{tau:cond2}) формулируется как $\partial_\mu\tau^\mu=0$, то есть 4-дивергенция векторного поля $\tau^\mu$ равна нулю.},  $*$ -- оператор Ходжа $* : \Lambda_k(M)\to \Lambda_{4-k}(M)$.

3)  На лоренцевом многообразии $(M,g)$ задана четверка гладких ортонормированных векторных полей $y^\mu_a$, называемая гладким полем реперов или тетрадой \cite{Bichteler1968}, занумерованная латинским индексом $a=0,1,2,3$ и
удовлетворяющая условиям
\begin{eqnarray}
&& y^\mu_a y^\nu_b\eta^{ab}=g^{\mu\nu},\quad \mu,\nu=0,1,2,3,\label{tetrada:cond1}\\
&& y^\mu_0 = \tau^\mu,\quad \mu=0,1,2,3.\label{tetrada:cond2}
\end{eqnarray}
Условие (\ref{tetrada:cond2}) означает, что тетрада $y^\mu_a$ {\em согласована} с векторным полем $\tau^\mu$.

Рассмотрим другую тетраду $\acute y^\mu_a$,  получающуюся из тетрады $y^\mu_a$ по формуле
\begin{equation}
 \acute y^\mu_a=r_a^b y^\mu_b,\label{yrab}
\end{equation}
 где матрица $R$  с элементами $r_a^b=r_a^b(x)$, $a,b=0,1,2,3$ есть гладкая функция от $x\in M$ со значениями в группе Ли матриц $SO_+^*(1,3)$ (см. (\ref{SOplusStar})).
В этом случае тетрада $\acute y^\mu_a$ удовлетворяет тем же условиям (\ref{tetrada:cond1}), (\ref{tetrada:cond2}) и можно говорить о классе тетрад, удовлетворяющих условиям (\ref{tetrada:cond1}), (\ref{tetrada:cond2}). Так как времениподобное векторное поле $y_0^\mu$ не меняется и совпадает с полем $\tau^\mu$, то этот класс будем называть {\em классом $\tau$-стабилизированных  тетрад}.
\medskip

Таким образом, в предлагаемой модели физическое пространство-время моделируется комплектом $(M,g,\tau,y)$, где комплект $(M,g)$ -- лоренцево многообразие; $\tau$ -- времениподобное нормированное векторное поле $\tau^\mu$, удовлетворяющее условиям
(\ref{tau:cond1}), (\ref{tau:cond2}); $y$ -- тетрада $y^\mu_a$, согласованная с векторным полем $\tau^\mu$ (класс тетрад, удовлетворяющих условиями (\ref{tetrada:cond1}), (\ref{tetrada:cond2})).

 Комплект $(M,g,\tau,y)$ будем обозначать рукописной буквой $\M$ и называть многообразием $\M$.
 \medskip

 \noindent{\bf  Замечание.} Стартовав с лоренцева многообразия $(M,g)$, дополнив его векторным полем $\tau^\mu$  и тетрадным полем $y^\mu_a$, мы пришли к многообразию $\M=(M,g,\tau,y)$. Возможен и другой путь, который следует подходу К.~Моллера (C.~M\o ller) к теории гравитации \cite{Moller}. А именно,
рассмотрим четырехмерное ориентируемое дифференцируемое многообразие $M$ с локальными  координатами $x^\mu$ и с заданной четверкой линейно независимых векторных полей $y^\mu_a$. Тогда, с помощью диагональной матрицы Минковского $\eta$ можно определить метрический тензор
$$
g^{\mu\nu}=y^\mu_a y^\nu_b\eta^{ab}
$$
лоренцевой сигнатуры, позволяющий поднимать/опускать тензорные (греческие) индексы, определяющий связность Леви-Чивиты и т.д. Соответственно, можно ввести векторное поле $\tau^\mu$, удовлетворяющее условиям  (\ref{tau:cond1}), (\ref{tau:cond2}). С этой точки зрения комплект $(M,g,\tau,y)$ фактически эквивалентен комплекту $(M,\tau,y)$ и эквивалентен комплекту $(M,y)$ (если мы рассматриваем класс тетрад $y^\mu_a$ для которого времениподобное векторное поле $y^\mu_0=\tau^\mu$ не меняется).

%%%%%%%%%%%%%%%%%%%%%%%%%%%%%%%%%%%%%%%%%%%%%%%%%%%%%%%%%%%%%%%%%%%%%%%%%%%%%%%%%%
\medskip

\noindent{\bf Тензорные поля со значениями в $\Mat(2,\C)$ на многообразии $\M$}.
В предлагаемой модели теории поля используются вещественные и комплексные тензорные поля на лоренцевом многообразии $(M,g)$ \cite{NovikovTaymanov}, а также тензорные поля со значениями в алгебре матриц $\Mat(2,\C)$.

Множество вещественных  тензорных полей типа $(r,s)$ обозначим через $\T^r_s$ (вместо $\T^r_0$, $\T^0_s$ пишем $\T^r$, $\T_s$), а комплексных тензорных полей через $\C\T^r_s$. Тензорное поле типа $(r,s)$ определяется в локальных координатах $x^\mu$ своими компонентами
$$
u^{\mu_1\ldots\mu_r}_{\nu_1\ldots\nu_s}\in\T^r_s,
$$
где индексы $\mu_1,\ldots,\mu_r,\nu_1,\ldots,\nu_s$ пробегают значения $0,1,2,3$. При этом компоненты тензорного поля
$$
u^{\mu_1\ldots\mu_r}_{\nu_1\ldots\nu_s}=u^{\mu_1\ldots\mu_r}_{\nu_1\ldots\nu_s}(x),\quad x\in M
$$
рассматриваются (локально) как гладкие функции $M\to\R$, либо $M\to\C$. При замене локальных координат многообразия $M$ компоненты тензорного поля преобразуются по стандартному правилу тензорного анализа \cite{NovikovTaymanov}.

Следуя \cite{AbramovAA}, наряду с выражением ``$u^{\mu_1\ldots\mu_r}_{\nu_1\ldots\nu_s}$ -- компоненты тензорного поля (тензора)'' допускается выражение ``$u^{\mu_1\ldots\mu_r}_{\nu_1\ldots\nu_s}$ -- тензорное поле (тензор)''; также будем писать
$u : u^{\mu_1\ldots\mu_r}_{\nu_1\ldots\nu_s}$,  $u\in\T^r_s$, $u^{\mu_1\ldots\mu_r}_{\nu_1\ldots\nu_s}\in\T^r_s$.
\medskip

\noindent{\em Сложение.}
Пусть $u : u^{\mu_1\ldots\mu_r}_{\nu_1\ldots\nu_s}$, $v : v^{\mu_1\ldots\mu_r}_{\nu_1\ldots\nu_s}$ -- два однотипных тензорных поля. Определена операция сложения тензорных полей $w=u+v$, если
\begin{equation}
w^{\mu_1\ldots\mu_r}_{\nu_1\ldots\nu_s}=u^{\mu_1\ldots\mu_r}_{\nu_1\ldots\nu_s}+v^{\mu_1\ldots\mu_r}_{\nu_1\ldots\nu_s},
\label{tensor:1}
\end{equation}
где $w$ -- тензорное поле того же типа.
\medskip

\noindent{\em Умножение.}
Пусть $u : u^{\mu_1\ldots\mu_r}_{\nu_1\ldots\nu_s}$, $v : v^{\alpha_1\ldots\alpha_k}_{\beta_1\ldots\beta_l}$ -- два тензорных поля. Определена операция умножения тензорных полей (тензорное произведение) $w=u\otimes v$, если
\begin{equation}
w^{\mu_1\ldots\mu_r\alpha_1\ldots\alpha_k}_{\nu_1\ldots\nu_s\beta_1\ldots\beta_l}=
u^{\mu_1\ldots\mu_r}_{\nu_1\ldots\nu_s} v^{\alpha_1\ldots\alpha_k}_{\beta_1\ldots\beta_l}.\label{tensor:2}
\end{equation}
где $w\in\T^{r+k}_{s+l}$.

Операция тензорного произведения тензорных полей обладает свойством ассоциативности
$(u\otimes v)\otimes w=u\otimes(v\otimes w)$.

Метрический тензор $g_{\mu\nu}$ лоренцева многообразия $(M,g)$ позволяет определить операции поднятия/опускания тензорных индексов у тензорных полей \cite{NovikovTaymanov}.

Для гладких тензорных полей лоренцева многообразия $(M,g)$ определена операция ковариантного дифференцирования (связность Леви-Чивиты) $\nabla_\mu : \T^r_s\to\T^r_{s+1}$ \cite{NovikovTaymanov}. Например, для векторного и ковекторного полей
$$
\nabla_\mu u^\nu =\partial_\mu u^\nu + \Gamma^\nu_{\mu\lambda}u^\lambda,\quad
\nabla_\mu u_\nu =\partial_\mu u_\nu - \Gamma^\lambda_{\mu\nu}u_\lambda,
$$
где символы Кристоффеля $\Gamma^\lambda_{\mu\nu}$ определяются метрическим тензором $g_{\mu\nu}$ \cite{NovikovTaymanov}.
Ковариантная производная от тензорного произведения тензорных полей удовлетворяет правилу Лейбница
$$
\nabla_\mu(u\otimes v)=(\nabla_\mu u)\otimes v + u\otimes(\nabla_\mu v).
$$

Наряду с обычными тензорными полями (со значениями в $\R$ и $\C$) на многообразии $\M$ будем рассматривать тензорные поля со значениями в $\Mat(2,\C)$ и, в частности, в  $\Herm(2)$, $\u(2)$, $\su(2)$, $\u(1)$ (то есть тензорные поля  компоненты которых $V^{\mu_1\ldots\mu_k}_{\nu_1\ldots\nu_l}$, являются матрицами второго порядка). В этом случае пишем $V^{\mu_1\ldots\mu_k}_{\nu_1,\ldots,\nu_l}\in\Mat(2,\C)\T^k_l$ ($V\in \Herm(2)\T^k_l$, либо $V\in \u(2)\T^k_l$, либо $V\in \su(2)\T^k_l$, либо $V\in \u(1)\T^k_l$).

%Если компоненты тензорного поля $u\in\CX\T^k_l$ принадлежат подпространству $\X_+$, или $\X_-$, или $i\X_+$, или $i\X_-$, то пишем $u\in\X_+\T^k_l$, $u\in\X_-\T^k_l$, $u\in i\X_+\T^k_l$, $u\in i\X_-\T^k_l$, соответственно.

%Если тензорное поле $u\in\C\otimes\X\T^k_l$ принадлежит некоторой алгебре Ли $L\subset\C\otimes\X$, то пишем $u\in L\T^k_l$.

Отметим, что обычному тензорному полю $u\in\T^k_l$ можно сопоставить тензорное поле со значениями в алгебре матриц $\Mat(2,\C)$ по правилу $u^{\mu_1\ldots\mu_k}_{\nu_1,\ldots,\nu_l}\to u^{\mu_1\ldots\mu_k}_{\nu_1,\ldots,\nu_l}e\in\Mat(2,\C)\T^k_l$, где $e$ -- единичная матрица второго порядка.
Тензорное поле $V^{\mu_1\ldots\mu_r}_{\nu_1\ldots\nu_s}$ со значениями в алгебре матриц $\Mat(2,\C)$ можно записать с помощью базиса Паули (\ref{Pauli1}) и четверки (нумеруется индексом $a=0,1,2,3$) обычных комплексных тензорных полей
\begin{equation}
V^{\mu_1\ldots\mu_r}_{\nu_1\ldots\nu_s}=v^{\mu_1\ldots\mu_r}_{\nu_1\ldots\nu_s a}\sigma^a\in\Mat(2,\C)\T^r_s,\label{formx}
\end{equation}
где
$$
v^{\mu_1\ldots\mu_r}_{\nu_1\ldots\nu_s a}\in\C\T^r_s,\quad a=0,1,2,3
$$
есть четверка тензорных полей, занумерованных латинским индексом $a$ и в правой части формулы (\ref{formx}) ведется суммирование по индексу $a$.

Компоненты тензорных полей со значениями в $\Mat(2,\C)$ будем, как правило, обозначать заглавными буквами.

Для тензорных полей со значениями в $\Mat(2,\C)$ определена операция сложения однотипных тензорных полей (\ref{tensor:1}) и определена операция (ассоциативного) тензорного произведения тензорных полей (\ref{tensor:2}).  При этом обращаем внимание на то, что в правой части формулы (\ref{tensor:2}) стоит обычное (матричное) произведение матриц
$$
U^{\mu_1\ldots\mu_r}_{\nu_1\ldots\nu_s}\quad\mbox{и}\quad V^{\alpha_1\ldots\alpha_k}_{\beta_1\ldots\beta_l}.
$$

Для гладких тензорных полей со значениями в $\Mat(2,\C)$ многообразия $\M$ определена операция ковариантного дифференцирования
$$
\nabla_\mu : \Mat(2,\C)\T^r_s\to\Mat(2,\C)\T^r_{s+1}
$$
с помощью формулы для $V^{\mu_1\ldots\mu_r}_{\nu_1\ldots\nu_s}=v^{\mu_1\ldots\mu_r}_{\nu_1\ldots\nu_s a}\sigma^a\in\Mat(2,\C)\T^r_{s}$
$$
\nabla_\alpha V^{\mu_1\ldots\mu_r}_{\nu_1\ldots\nu_s}:=(\nabla_\alpha v^{\mu_1\ldots\mu_r}_{\nu_1\ldots\nu_s a})\sigma^a\in\Mat(2,\C)\T^r_{s+1}.
$$
При этом ковариантная производная от тензорного произведения тензорных полей со значениями в $\Mat(2,\C)$ удовлетворяет правилу Лейбница
$$
\nabla_\mu(U\otimes V)=(\nabla_\mu U)\otimes V + U\otimes(\nabla_\mu V).
$$
%%%%%%%%%%%%%%%%%%%%%%%%%%%
%%%%%%%%%%%%%%%%%%%%%%%%%%%%%%%%

\section{Векторное поле $K^\mu\in\Herm(2)\T^1$}

Для того, чтобы написать уравнения поля нам потребуется векторное поле, обозначаемое $K^\mu$, со значениями  в пространстве эрмитовых матриц второго порядка $\Herm(2)$, удовлетворяющее следующим двум условиями:
\begin{eqnarray}
&& K^\mu\tilde K^\nu + K^\nu\tilde K^\mu=2 g^{\mu\nu}e,\quad\mu,\nu=0,1,2,3,\label{cond:kmu1}\\
&& \nabla_\mu\pi_+(K^\mu)=0.\label{cond:kmu2}
\end{eqnarray}

Вектор (векторное поле) $K^\mu\in\Herm(2)\T^1$, удовлетворяющий условиям (\ref{cond:kmu1}) и (\ref{cond:kmu2}), будем называть {\em генвектором}.

%(????????????а надо ли это название, здесь $K^\mu$ уже не есть генераторы алг Клиффорда).

Набор из четырех матриц $K^\mu$ можно рассматривать как базис в вещественном векторном пространстве $\Herm(2)$, либо как базис в комплексном векторном пространстве $\Mat(2,\C)$.

Если возьмем скалярное поле $S=S(x)\in \SU(2)\subset\Mat(2,\C)$ и определим другое векторное поле
$$
\acute K^\mu = S^{-1}K^\mu S\in\Herm(2)\T^1,
$$
то, как легко убедиться, векторное поле $\acute K^\mu$ тоже удовлетворяет условиям (\ref{cond:kmu1}),(\ref{cond:kmu2}).

Векторное поле $K^\mu$ можно опеределить с помощью тетрады $y^\mu_a$, удовлетворяющей условиям (\ref{tetrada:cond1}), (\ref{tetrada:cond2})  и с помощью базиса Паули $\sigma^a$
\begin{equation}
K^\mu =y^\mu_a \sigma^a\in\Herm(2)\T^1.\label{Kmu:y}
\end{equation}
Отметим, что
\begin{equation}
K^\mu=\pi_+(K^\mu) + \pi_-(K^\mu) = K_+^\mu + K_-^\mu\label{Kmu:dec}
\end{equation}
и
\begin{equation}
K_+^\mu=\tau^\mu e.\label{Kmu:dec1}
\end{equation}

Легко убедиться, что компоненты этого вектора удовлетворяют соотношениям (\ref{cond:kmu1}), а
условие на тетраду (\ref{tetrada:cond2}) и условие $\nabla_\mu\tau^\mu=0$  дает для вектора $K^\mu$ соотношение (\ref{cond:kmu2}). То есть векторное поле $K^\mu$, заданное формулой (\ref{Kmu:y}),  является генвектором.

Рассмотрим другую тетраду $\acute y^\mu_a$,  получающуюся из тетрады $y^\mu_a$ с помощью формулы (\ref{yrab}).
В этом случае тетрада $\acute y^\mu_a$ удовлетворяет тем же условиям (\ref{tetrada:cond1}), (\ref{tetrada:cond2}) и определяет генвектор
$$
\acute K^\mu=\acute y^\mu_a\sigma^a.
$$
\begin{sloppypar}
Таким образом, класс генвекторов можно связать с введенным классом $\tau$-стабилизированных  тетрад (удовлетворяющих (\ref{tetrada:cond1}), (\ref{tetrada:cond2})).
\end{sloppypar}

Класс генвекторов можно описать и с другой точки зрения. А именно,  рассматривая функции $M\to\Mat(2,\C)$,
можно ввести базис (класс базисов), зависящий от $x\in M$ и выражающийся через базис Паули по формуле
\begin{equation}
\acute\sigma_0 =\sigma_0,\quad \acute\sigma_k=p_k^l\sigma_l,\quad k=1,2,3\label{sigma:sigmaprime}
%\rho_a=S^{-1}\sigma_a S,\quad a=0,1,2,3,\label{rho:sigma0}
\end{equation}
где в правой части стоит сумма по $l=1,2,3$ и матрица $P=\|p_k^l\|$ (функция от $x\in M$ со значениями в $SO(3)$) принадлежит группе $SO(3)$. В этом случае класс генвекторов можно определить с помощью одной тетрады $y^\mu_a$ (удовлетворяющей (\ref{tetrada:cond1}), (\ref{tetrada:cond2})) и класса базисов $\acute\sigma^a$, состоящих из эрмитовых матриц
$$
\acute K^\mu = y^\mu_a\acute\sigma^a.
$$
Если рассмотреть равенства (\ref{sigma:sigmaprime}) в фиксированной точке $\st{\circ}{x}\in M$, то можно воспользоваться теоремой о двойном накрытии групп \cite{MarShir2021}
$$
\SU(2)/\{\pm1\}\simeq \SO(3),
$$
в соответствии с которой для каждой матрицы $P\in \SO(3)$ существует единственная пара матриц $\pm S$ из $\SU(2)$ такая, что
\begin{equation}
S^{-1}\sigma_k S = p_k^l\sigma_l,\quad k=1,2,3,\label{SU2SO3:1}
\end{equation}
и наоборот, для каждой пары матриц $\pm S$ из $\SU(2)$ существует единственная матрица $P\in \SO(3)$ такая, что выполнены соотношения (\ref{SU2SO3:1}).
Сравнивая формулы (\ref{sigma:sigmaprime}) и (\ref{SU2SO3:1}), получим
\begin{equation}
\acute\sigma^a = p_b^a\sigma^b=S^{-1}\sigma^a S,\quad a=0,1,2,3,\label{SU2SO3:1x}
\end{equation}
где $p^0_0=1$, $p^0_k=p^k_0=0$, $k=1,2,3$.
При попытке применить формулу (\ref{SU2SO3:1x}) к функциям от $x\in M$ возникают трудности \cite{Azer} из-за которых удается получить результат лишь в малой окрестности точки $\st{\circ}{x}\in M$.

%%%%%%%%%%%%%%%%%%%%%%%%%%%%%%%%%%%%%

Некоторые свойства матриц  матриц $K^\mu$ сформулируем в виде теоремы.

\begin{theorem}\label{theorem1}
Матрицы  $K^\mu=y^\mu_a\sigma^a$, кроме равенств (\ref{cond:kmu1}), (\ref{cond:kmu2}), удовлетворяют следующим равенствам ($\forall A\in\Mat(2,\C)\T^r_s$):
%???????????????формулы еще не проверены
\begin{eqnarray}
K^\mu + \tilde K^\mu &=& 2\tau^\mu e,\\
K^\mu\tilde K_\mu &=& K_\mu\tilde K^\mu=
\tilde K^\mu K_\mu = \tilde K_\mu K^\mu
 = 4 e,\label{K:prop:0}\\
\frac{1}{2}\tr(K^\mu\tilde K_\nu) &=& \frac{1}{2}\tr(\tilde K^\mu K_\nu)=\delta^\mu_\nu,\label{K:prop:1}\\
\frac{1}{2}K^\mu A K_\mu &=&\frac{1}{2}\tilde K^\mu A\tilde K_\mu=-\tilde A=-A_+ + A_- ,\label{K:prop:2}\\
\frac{1}{2}\tilde K^\mu A K_\mu &=&\frac{1}{2}K^\mu A\tilde K_\mu=(\tr\,A)e=2 A_+,\label{K:prop:3}\\
y^\mu_b &=& \frac{1}{2}\tr(K^\mu\tilde\sigma_b),\label{K:prop:4}\\
(\nabla_\mu y^\lambda_a)y^a_\nu &=& \frac{1}{4}\tr((\nabla_\mu K^\lambda)\tilde\sigma_a)\tr(K_\nu\tilde\sigma^a).\label{K:prop:5}
\end{eqnarray}
\end{theorem}

Для любого тензора (тензорного поля) со значениями в матрицах $A^{\mu_1\ldots\mu_r}_{\nu_1\ldots\nu_s}\in\Mat(2,\C)\T^r_s$ его разложение по базису $K^\beta$ дается формулой
\begin{equation}
A^{\mu_1\ldots\mu_r}_{\nu_1\ldots\nu_s} = a^{\mu_1\ldots\mu_r}_{\nu_1\ldots\nu_s\beta} K^\beta,\quad
  a^{\mu_1\ldots\mu_r}_{\nu_1\ldots\nu_s\beta}= \frac{1}{2}\tr(A^{\mu_1\ldots\mu_r}_{\nu_1\ldots\nu_s}\tilde K_\beta).\label{AaK}
\end{equation}
В частности, для матрицы $A\in\Mat(2,\C)$ (скалярное поле со значениями в алгебре матриц) имеем разложение по базису $K^\beta$
\begin{equation}
A = a_{\beta} K^\beta,\quad
  a_{\beta}= \frac{1}{2}\tr(A\tilde K_\beta).\label{AaK0}
\end{equation}

%%%%%%%%%%%%%%%%%%%%%%%%
\medskip

\noindent{\bf Вектор $K^\mu$ как квадратный корень из метрического тензора $g^{\mu\nu}$.}
%???????????????Следующее утверждение нуждается в проверке.
Рассмотрим квадратичную форму, ассоциированную с метрическим тензором $g^{\mu\nu}$ лоренцева многообразия $(M,g)$
$$
g^{\mu\nu}\xi_\mu\xi_\nu,
$$
где $\xi_\mu$ -- коммутирующие символы ($\mu=0,1,2,3$). Теперь возьмем две линейные формы, ассоциированные с векторами $K^\mu$ и $\tilde K^\mu$. Тогда, с помощью условия (\ref{cond:kmu1}) легко убедиться в справедливости равенства
\begin{equation}
g^{\mu\nu}\xi_\mu\xi_\nu\, e = (K^\mu\xi_\mu)(\tilde K^\nu\xi_\nu)= (\tilde K^\mu\xi_\mu)(K^\nu\xi_\nu),\quad\forall x\in M.
\label{gKK}
\end{equation}
Тем самым, квадратичную форму с метрическим тензором мы представили в виде произведения двух линейных форм с векторами $K^\mu$ и $\tilde K^\mu$.

Выписанная формула (\ref{gKK}) указывает в каком смысле векторное поле $K^\mu$ можно интерпретировать как квадратный корень из метрического тензора $g^{\mu\nu}$.

%%%%%%%%%%%%%%%%%%%%%%%%%%%%%%%%%%%%%%%%%%%%

\section{Система уравнений Янга-Миллса\\ на лоренцевом многообразии}

Дальше будем рассматривать системы дифференциальных уравнений, которые в модели используются для описания физических полей (элементарных частиц и их взаимодействий).

Пусть $G$ -- некоторая группа Ли унитарных матриц (с матричным умножением), и $L$ -- вещественная алгебра Ли группы Ли $G$ (скобка Ли отождествляется с коммутатором матриц). Рассмотрим систему уравнений Янга-Миллса\footnote{В этой статье мы рассматриваем несколько калибровочных полей и, для удобства, обозначаем потенциал и напряженность поля Янга-Миллса одно и той же буквой, но с разным числом индексов (с одним индексом -- потенциал поля, а с двумя -- напряженность поля).
%Например $A_\mu$ -- потенциал поля Янга-Миллса, а $A_{\mu\nu}$ -- напряженность поля Янга-Миллса.
}  (1954) на  $\M$
\begin{equation}
\nabla_\mu A_\nu-\nabla_\nu A_\mu - [A_\mu, A_\nu] = A_{\mu\nu},\quad
\nabla_\mu A^{\mu\nu}- [A_\mu, A^{\mu\nu}] = J^\nu,\label{YM:aa}
\end{equation}
 где $A_\mu\in L\T_1$, $A_{\mu\nu}\in L\T_2$, $J^\nu\in L\T^1$ и эти тензорные поля зависят от  $x\in M$. Система уравнений Янга-Миллса инвариантна относительно калибровочного преобразования (симметрии)
\begin{equation}
A_\mu\to S^{-1}A_\mu S - S^{-1}\nabla_\mu S,\quad A_{\mu\nu}\to S^{-1}A_{\mu\nu}S,\quad
J^\nu \to S^{-1}J^\nu S,\label{gauge:sym:0}
\end{equation}
где $S=S(x)\in G$. Известно, что система уравнений Янга-Миллса
 (\ref{YM:aa}) имеет следствие
\begin{equation}
\nabla_\nu J^\nu - [A_\nu, J^\nu] =0.\label{YM:conseq}
\end{equation}
Пара $(A_\mu,A_{\mu\nu})$, удовлетворяющая системе уравнений (\ref{YM:aa}), называется
{\em полем Янга-Миллса}; $A_\mu$ -- потенциал поля Янга-Миллса и $A_{\mu\nu}$ -- напряженность поля Янга-Миллса.
%%%%%%%%%%%%%%%%%%%%%%%%%%%%%%%%%%%%%%%%%%%%%%%%%%%%%%

\section{Примитивное полевое уравнение\\ в алгебре матриц $\Mat(2,\C)$}

Если  $\gamma^\nu=\gamma^\nu(x)$ -- обобщенные $4\times4$-матрицы Дирака, заданные на лоренцевом многообразии $M$ и удовлетворяющие соотношениям ($\Id_4$ -- единичная матрица четвертого порядка)
$$
\gamma^\mu\gamma^\nu+\gamma^\nu\gamma^\mu=2 g^{\mu\nu}\Id_4,
$$
то {\em примитивное полевое уравнение} ($\nabla_\mu$ -- ковариантная производная - связность Леви-Чивиты)
$$
\nabla_\mu\gamma^\nu - [\Gamma_\mu,\gamma^\nu]=0,
$$
позволяет найти выражение для спиновой связности ($\Gamma_\mu$ -- ковектор со значениями в $\Mat(4,\C)$)
$$
\Gamma_\mu = \frac{1}{4}(\nabla_\mu\gamma^\nu)\gamma_\nu,
$$
которая используется в записи уравнения Дирака для электрона на лоренцевом многообразии с метрическим тензором $g_{\mu\nu}$
(\cite{Mitsk}, пар. 4.5).

В \cite{MarShirROMP}, \cite{Marchuk2018} примитивное полевое уравнение было перенесено на алгебры Клиффорда и найден новый класс решений этого уравнения.

В этом параграфе будет дано дальнейшее развитие концепции примитивного полевого уравнения в двух направлениях. Во-первых, уравнение будет перенесено на матрицы из $\Mat(2,\C)$. Во-вторых, в уравнение будет добавлено еще одно слагаемое (правая часть), которое даст возможность использовать это уравнение для описания гравитационного поля в рамках рассматриваемой модели.

На многообразии $(M,g,\tau,y)$ рассматриваем
 $K^\mu\in\Herm(2)\T^1$ -- генвектор (векторное поле), $C_\lambda\in\su(2)\T_1$ -- ковекторное поле со значениями в алгебре Ли $\su(2)$, $\varkappa_{\lambda\mu}{}^\nu\in\T^1_2$ -- вещественное тензорное поле, удовлетворяющее условиям
\begin{equation}
 \varkappa_{\lambda\mu}{}^\lambda=0,\quad
 \varkappa_{\lambda\mu\nu} = -\varkappa_{\lambda\nu\mu}.\label{varkappa}
 \end{equation}

  Уравнение
 \begin{equation}
 \nabla_\lambda K^\nu - [C_\lambda, K^\nu]= \varkappa_{\lambda\mu}{}^\nu K^\mu \label{prim:eq:mat}
 \end{equation}
 будем называть {\em примитивным полевым уравнением}.\footnote{Уравнение, называемое  {\em примитивным полевым уравнением} содержащим тензорные величины в пространстве Минковского со значениями в алгебре Клиффорда, рассматривалось в ряде работ автора (см., например, \cite{Marchuk2018} и др.).  В настоящей работе примитивное полевое уравнение содержит тензорные величины со значениями в алгебре матриц $\Mat(2,\C)$ и рассматривается на многообразии  $(M,g,\tau,y)$.}

Из первого условия (\ref{varkappa}) видим, что
следствием примитивного полевого уравнения (\ref{prim:eq:mat}) будет равенство
\begin{equation}
\nabla_\lambda K^\lambda - [C_\lambda, K^\lambda]=0.\label{prim:eq:con}
 \end{equation}

Легко проверить, что уравнение (\ref{prim:eq:mat}) является инвариантным относительно калибровочного преобразования с группой симметрии $\SU(2)$
\begin{equation}
K^\nu \to \acute K^\nu=S^{-1}K^\nu S,\quad
C_\lambda \to \acute C_\lambda = S^{-1} C_\lambda S - S^{-1}\partial_\lambda S,\quad
\varkappa_{\lambda\mu}{}^\nu\to \varkappa_{\lambda\mu}{}^\nu,
\end{equation}
где $S=S(x)\in\SU(2)$, $x\in M$.

Подействуем операторами проектирования $\pi_\pm$ на обе части уравнения (\ref{prim:eq:mat}) и воспользуемся обозначениями
$$
\tau^\nu e = \pi_+(K^\nu),\quad L^\nu = \pi_-(K^\nu).
$$
Видим, что уравнение (\ref{prim:eq:mat}) распадается на два уравнения
\begin{eqnarray}
\nabla_\lambda\tau^\nu &=& \varkappa_{\lambda\mu}{}^\nu \tau^\mu,\label{1st:eq}\\
\nabla_\lambda L^\nu -[C_\lambda, L^\nu] &=& \varkappa_{\lambda\mu}{}^\nu L^\mu.\label{2nd:eq}
\end{eqnarray}

\begin{theorem}\label{theorem2}
Из уравнения (\ref{2nd:eq}) можно выразить ковекторное поле
$C_\lambda$ через векторное поле $L^\nu$ и тензорное поле $\varkappa_{\lambda\mu}{}^\nu$
\begin{equation}
C_\lambda = -\frac{1}{4}\big((\nabla_\lambda L^\nu)L_\nu -
\varkappa_{\lambda\mu\nu} L^\mu L^\nu \big).
\label{CmuViaKnu}
\end{equation}
\end{theorem}
 \begin{proof}.
 Надо обе части уравнения (\ref{prim:eq:mat}) справа умножить на $L_\nu$ (и свернуть по индексу $\nu$). С помощью равенств
$$
L^\nu L_\nu = -3 e,\quad L^\nu C_\lambda L_\nu = C_\lambda-4\pi_+(C_\lambda)=C_\lambda
$$
результат может быть записан в виде (\ref{CmuViaKnu}).
 При этом отметим, что второе равенство в (\ref{varkappa}) обеспечивает принадлежность правой части к $\su(2)\T_1$.
\end{proof}

%%%%%%%%%%%%%%%%%%%

\section{Основная система уравнений с $\SU(2)$ и $\U(2)$ калибровочной симметрией}

%Начиная с этой страницы надо заменить $\partial_\mu\to\nabla_\mu$ !!!!!!!!!!!!!!!!!!!!!!!

Пусть векторное поле $K^\mu\in\Herm(2)\T^1$ удовлетворяет условиям (\ref{cond:kmu1}), (\ref{cond:kmu2}). И пусть заданы ковекторные поля $A_\mu \in \u(2)\T_1$, $C_\mu\in \su(2)\T_1$.

Модельное уравнение Дирака-Ланцоша\footnote{Это уравнение можно рассматривать как существенную модификацию уравнения Дирака-Ланцоша (\cite{LanczosDirac}, формула (63)), которое было записано в технике бикватернионов \cite{Kravchenko}. Отметим следующие отличия уравнения (\ref{mod:Lanczos:Z}) от уравнения Дирака-Ланцоша:
\begin{itemize}
\item Уравнение Дирака-Ланцоша является спинорным, а уравнение
(\ref{mod:Lanczos:Z}) является тензорным.
\item В отличие от уравнения Дирака-Ланцоша, уравнение (\ref{mod:Lanczos:Z}) обладает двумя неабелевыми калибровочными симметриями относительно унитарных групп Ли $\SU(2)$ и $\U(2)$.
\item Уравнение Дирака-Ланцоша рассматривается  в пространстве Минковского, а уравнение (\ref{mod:Lanczos:Z}) рассматривается на псевдоримановом многообразии лоренцевой сигнатуры.
\end{itemize}
} (с нулевой массой\footnote{В стандартное уравнение Дирака для электрона $i\gamma^\mu(\partial_\mu\psi + i a_\mu\psi)- m\psi=0$ входит слагаемое с массой электрона $m$. Вместе с тем, при построении теории электрослабых взаимодействий слагаемое с массой $m$ нарушает необходимую калибровочную симметрию левоспиральной компоненты волновой функции по отношению к калибровочной группе $\SU(2)$. Как известно, для решения этой проблемы в электрослабой теории предложено убрать слагаемое с массой в соответствующем уравнении Дирака и считать, что первоначально частица является безмассовой, а потом она приобретает массу путем юкавского взаимодействия с полем Хиггса.}) на многообразии $\M$ для волновой функции $\Psi=\Psi(x)\in\Mat(2,\C)$ имеет вид
\begin{equation}
i K^\mu(\nabla_\mu\Psi -C_\mu\Psi +\Psi A_\mu)=0.\label{mod:Lanczos:Z}
\end{equation}

\begin{theorem}\label{theorem3}
Пусть на многообразии $\M$ даны (a) векторное поле $K^\mu\in\Herm(2)\T^1$, удовлетворяющее условиям (\ref{cond:kmu1}), (\ref{cond:kmu2}); (b) ковекторные поля $A_\mu \in \u(2)\T_1$, $C_\mu\in \su(2)\T_1$; (c) скалярное поле $\Psi=\Psi(x)\in\Mat(2,\C)$. И пусть эти поля удовлетворяют уравнению (\ref{mod:Lanczos:Z}). Тогда эти поля удовлетворяют равенству
\begin{equation}
(\nabla_\mu(\Psi^\dagger i K^\mu\Psi) - [A_\mu, \Psi^\dagger i K^\mu\Psi])
- \Psi^\dagger(\nabla_\mu\pi_-(i K^\mu) - [C_\mu, \pi_-(i K^\mu)])\Psi=0.\label{cons:kpsi:Z0}
\end{equation}
\end{theorem}
\begin{proof}.
 Уравнение (\ref{mod:Lanczos:Z}) умножим слева на $\Psi^\dagger$ и возьмем эрмитово сопряжение  от получившегося равенства. С учетом соотношений
$$
(i K^\mu)^\dagger=-i K^\mu,\quad C_\mu^\dagger=-C_\mu,\quad (A_\mu)^\dagger=-A_\mu,
$$
будем иметь
\begin{eqnarray}
&& \Psi^\dagger i K^\mu(\nabla_\mu\Psi -C_\mu\Psi +\Psi A_\mu)=0,\label{psidagpsi1:Z}\\
&& (\nabla_\mu\Psi^\dagger+\Psi^\dagger C_\mu-A_\mu\Psi^\dagger)(-i K^\mu)\Psi=0.\label{psidagpsi2:Z}
\end{eqnarray}
Разность левых частей равенств (\ref{psidagpsi1:Z}) и (\ref{psidagpsi2:Z}) можно записать в виде
\begin{equation}
(\nabla_\mu(\Psi^\dagger i K^\mu\Psi) - [A_\mu, \Psi^\dagger i K^\mu\Psi])
- \Psi^\dagger(\nabla_\mu(i K^\mu) - [C_\mu, i K^\mu])\Psi=0.\label{cons:kpsi:Z}
\end{equation}
Отметим, что
\begin{eqnarray}
\nabla_\mu(\Psi^\dagger i K^\mu\Psi) - [A_\mu, \Psi^\dagger i K^\mu\Psi] &=&
(\nabla_\mu\pi_+(\Psi^\dagger i K^\mu\Psi))\nonumber\\
&&+ (\nabla_\mu\pi_-(\Psi^\dagger i K^\mu\Psi) - [\pi_-(A_\mu), \pi_-(\Psi^\dagger i K^\mu\Psi)]),\nonumber\\
\nabla_\mu(i K^\mu) - [C_\mu, i K^\mu] &=&
(\nabla_\mu\pi_+( i K^\mu))\nonumber\\
&&+ (\nabla_\mu\pi_-(i K^\mu) - [C_\mu, \pi_-(i K^\mu)])\label{ident55}
\end{eqnarray}
и напомним, что $\nabla_\mu\pi_+( i K^\mu)=0$ в силу (\ref{cond:kmu2}). Подставив  (\ref{ident55}) в равенство (\ref{cons:kpsi:Z}), получим (\ref{cons:kpsi:Z0}).
Теорема доказана.
\end{proof}

\begin{theorem}\label{theorem4}
Уравнение (\ref{mod:Lanczos:Z}) имеет следующие калибровочные симметрии:

I. Пусть $V=V(x)$ есть произвольная гладкая функция $M\to \U(2)$. Тогда уравнение (\ref{mod:Lanczos:Z}) является инвариантным относительно калибровочного преобразования
\begin{eqnarray}
&& \Psi \to\acute\Psi=\Psi V,\nonumber\\
&& K^\mu \to\acute K^\mu=K^\mu,\label{gauge:Lanczos1:Z}\\
&& C_\mu \to\acute C_\mu=C_\mu,\nonumber\\
&& A_\mu\to \acute A_\mu=V^{-1}A_\mu V- V^{-1}\nabla_\mu V.\nonumber
\end{eqnarray}

II. Пусть $S : M\to \SU(2)$ есть произвольная гладкая функция от $x\in M$ со значениями в группе Ли $\SU(2)$.
Тогда уравнение (\ref{mod:Lanczos:Z}) является инвариантным относительно калибровочного преобразования
\begin{eqnarray}
&& \Psi \to\acute\Psi=S^{-1}\Psi S,\nonumber\\
&& K^\mu \to\acute K^\mu=S^{-1}K^\mu S,\label{gauge:Lanczos2:Z}\\
&& C_\mu \to\acute C_\mu=S^{-1}C_\mu S -S^{-1}\nabla_\mu S,\nonumber\\
&& A_\mu\to \acute A_\mu= S^{-1}A_\mu S - S^{-1}\nabla_\mu S.\nonumber
\end{eqnarray}

III. Пусть $S : M\to \SU(2)$ есть произвольная гладкая функция от $x\in M$ со значениями в группе Ли $\SU(2)$. Возьмем суперпозицию преобразований (I) и (II) при $V=S^{-1}$.
Тогда уравнение (\ref{mod:Lanczos:Z}) является инвариантным относительно калибровочного преобразования
\begin{eqnarray}
&& \Psi \to\acute\Psi=S^{-1}\Psi,\nonumber\\
&& K^\mu \to\acute K^\mu=S^{-1}K^\mu S,\label{gauge:Lanczos3:Z}\\
&& C_\mu \to\acute C_\mu=S^{-1}C_\mu S -S^{-1}\nabla_\mu S,\nonumber\\
&& A_\mu\to \acute A_\mu= A_\mu.\nonumber
\end{eqnarray}
\end{theorem}

\begin{proof}. Доказательство теоремы проводится подстановкой штрихованных величин из формул (\ref{gauge:Lanczos1:Z}), (\ref{gauge:Lanczos2:Z}), (\ref{gauge:Lanczos3:Z}) в уравнение
(\ref{mod:Lanczos:Z}) и проверкой того, что штрихованные переменные удовлетворяют тому же уравнению.
\end{proof}
\medskip

\noindent{\bf Гамильтонова форма модельного уравнения Дирака-Ланцоша.}
Дифференциальный оператор $K^\mu\nabla_\mu$ с помощью формул (\ref{Kmu:dec}), (\ref{Kmu:dec1}) запишем в виде
$$
K^\mu\nabla_\mu = \tau^\mu\nabla_\mu +  K^\mu_-\nabla_\mu,
$$
 и обозначим $\nabla_{(\tau)} := \tau^\mu\nabla_\mu$ -- ковариантная производная в направлении касательного векторного поля $\tau^\mu$. Следовательно, модельное уравнение Дирака-Ланцоша
(\ref{mod:Lanczos:Z}) переписывается в виде, который можно назвать гамильтоновой формой модельного  уравнения Дирака-Ланцоша
\begin{equation}
i\nabla_{(\tau)}\Psi = H\Psi,\label{Hamil:eq}
\end{equation}
где
$$
H\Psi := -i K^\mu_-\nabla_\mu\Psi +
i K^\mu(C_\mu\Psi - \Psi A_\mu)
$$
и роль производной по времени играет $\nabla_{(\tau)}$ -- ковариантная производная в направлении времениподобного касательного векторного поля $\tau^\mu$. Вид уравнения (\ref{Hamil:eq}) указывает на возможность разработки релятивистского гамильтонова формализма для рассматриваемой модели теории поля (как развитие подхода \cite{KozlovNikishin}).

%%%%%%%%%%%%%%%%%%%%%%%%%%%%%%%%%%%%%%%%%

%%%%%%%%%%%%%%%%%%%%%%%%%%%%%%%%%%%%%%%%%%

\section{Уравнения для полей $A_\mu$, $C_\mu$, $K^\mu$}
\medskip

%Март 23, 2022, война в Украине
\medskip

Выписанное уравнение (\ref{mod:Lanczos:Z}) мы рассматриваем как уравнение для волновой функции фермиона $\Psi=\Psi(x)\in\Mat(2,\C)$ и считаем, что входящие в это уравнение величины $A_\mu$, $C_\mu$, $K^\mu$ известны (заданы) и удовлетворяют следующим  условиям: векторное поле $K^\mu\in\Herm(2)\T^1$ удовлетворяет условиям (\ref{cond:kmu1}), (\ref{cond:kmu2}) и ковекторные поля $A_\mu \in \u(2)\T_1$, $C_\mu\in \su(2)\T_1$. Для дальнейшего развития модели мы дополним уравнение (\ref{mod:Lanczos:Z}) некоторыми уравнениями для величин $A_\mu$, $C_\mu$, $K^\mu$ и посмотрим на получающуюся систему уравнений.

Начнем с того, что дополним уравнение (\ref{mod:Lanczos:Z}) уравнениями Янга-Миллса для потенциала $A_\mu \in \u(2)\T_1$. Полагаем, что в получившейся системе уравнений $\Psi$, $A_\mu$ -- неизвестные, а $C_\mu$, $K^\mu$ -- известные
 \begin{eqnarray}
&& i K^\mu(\nabla_\mu\Psi -C_\mu\Psi +\Psi A_\mu)=0,\label{main:eq1:Z0}\\
&& \nabla_\mu A_\nu-\nabla_\nu A_\mu-[A_\mu,A_\nu] = A_{\mu\nu},\label{main:eq4:Z0}\\
&& \nabla_\mu A^{\mu\nu}-[A_\mu,A^{\mu\nu}]=\Psi^\dagger i K^\nu\Psi.\label{main:eq5:Z0}
\end{eqnarray}
Эта система уравнений имеет те же калибровочные симметрии (\ref{gauge:Lanczos1:Z}), (\ref{gauge:Lanczos2:Z}), (\ref{gauge:Lanczos3:Z}), что и уравнение (\ref{mod:Lanczos:Z}).

Теперь надо включить уравнения для $C_\mu$, $K^\mu$. Что это за уравнения? Отметим, что равенство (\ref{cons:kpsi:Z}) содержит подсказку -- следствием новых уравнений для  $C_\mu$, $K^\mu$ должно быть равенство
\begin{equation}
\nabla_\mu K^\mu  - [C_\mu, K^\mu]=0.\label{cons:CK}
\end{equation}
Вижу три возможных варианта уравнений для  $C_\mu$, $K^\mu$.

Во-первых, примитивное полевое уравнение
\begin{equation}
\nabla_\mu K^\nu - [C_\mu, K^\nu]=\alpha(4\,\tau_\mu\tau^\nu-\delta^\nu_\mu)\Psi\Psi^\dagger,
\label{primitive:fe}
\end{equation}
в которое кроме волновой функции $\Psi$ входит вещественное скалярное поле $\alpha=\alpha(x)$. В записи примитивного полевого уравнения (\ref{prim:eq:mat}) в правой части мы использовали выражение $\varkappa_{\lambda\mu}{}^\nu K^\mu$. Если матрицу (скалярное поле со значениями  в матрицах) $\Psi\Psi^\dagger$ разложить по базису $K^\epsilon$ (по формуле
(\ref{AaK}))
$$
\Psi\Psi^\dagger = \phi_\mu K^\mu,\quad
\phi_\mu = \frac{1}{2}\tr(\Psi\Psi^\dagger\tilde K_\mu),
$$
то будем иметь
$$
\varkappa_{\lambda\mu}{}^\nu = \alpha(4\tau^\nu\tau_\lambda-\delta^\nu_\lambda)\phi_\mu.
$$

Во-вторых, уравнения Янга-Миллса
\begin{eqnarray}
&& \nabla_\mu C_\nu-\nabla_\nu C_\mu-[C_\mu,C_\nu] = C_{\mu\nu},\label{YM:A1}\\
&& \nabla_\mu C^{\mu\nu}-[C_\mu,C^{\mu\nu}]=\pi_-(i K^\nu)=i L^\nu.\label{YM:A2}
\end{eqnarray}

И, в-третьих, система уравнений состоящая из всех уравнений (\ref{YM:A1}), (\ref{YM:A2}), (\ref{primitive:fe})
\begin{eqnarray}
&& \nabla_\mu C_\nu-\nabla_\nu C_\mu-[C_\mu,C_\nu] = C_{\mu\nu},\label{YM:A1x}\\
&& \nabla_\mu C^{\mu\nu}-[C_\mu,C^{\mu\nu}]=\pi_-(i K^\nu),\label{YM:A2x}\\
&&\nabla_\mu(i K^\nu) - [C_\mu, i K^\nu]=i \alpha(4\,\tau_\mu\tau^\nu-\delta^\nu_\mu)\Psi\Psi^\dagger.
\label{primitive:fex}
\end{eqnarray}

Напомним, что ковекторное поле $C_\mu\in\su(2)\T_1$ мы интерпретировали как янг-миллсовский потенциал гравитационного поля, а векторное поле  $K^\nu\in\Herm(2)\T^1$, в силу формулы (\ref{gKK}), мы интерпретировали  как квадратный корень из метрического тензора. Поэтому уравнения (\ref{YM:A1}), (\ref{YM:A2}), (\ref{primitive:fe}) можно рассматривать как связь между гравитационным полем (с потенциалом $C_\mu$ и напряженностью $C_{\mu\nu})$, волновой функцией фермиона $\Psi$ и метрикой многообразия  (заданной квадратным корнем $K^\mu$ из метрического тензора $g^{\mu\nu}$).
%При обсуждении примитивного полевого уравнения (\ref{prim:eq:mat}) было указано, что это уравнение позволяет выразить ковекторное поле $C_\mu\in\su(2)\T_1$ через векторное поле $K^\nu\in\Herm(2)\T^1$ по формуле (\ref{CmuViaKnu}). Подставив выражения для $C_\mu$ в уравнения Янга-Миллса (\ref{main:eq2:Z}), (\ref{main:eq3:Z}), можно получить уравнение третьего порядка для векторного поля $K^\nu$.  И это уравнение  является калибровочно инвариантным относительно преобразования (\ref{gauge:Lanczos2:Z}) с калибровочной группой $\SU(2)$. Поэтому это уравнение третьего порядка (для $K^\mu$) можно рассматривать как уравнение на метрику многообразия.

Какой из указанных трех вариантов систем уравнений для $C_\mu$, $K^\mu$ окажется предпочтительным, мы надеемся определить в результате дальнейших исследований. А пока, в качестве основной системы уравнений, будем рассматривать систему уравнений
\begin{eqnarray}
&& i K^\mu(\nabla_\mu\Psi -C_\mu\Psi +\Psi A_\mu)=0,\label{main:eq1:Z}\\
&& \nabla_\mu A_\nu-\nabla_\nu A_\mu-[A_\mu,A_\nu] = A_{\mu\nu},\label{main:eq4:Z}\\
&& \nabla_\mu A^{\mu\nu}-[A_\mu,A^{\mu\nu}]=\Psi^\dagger i K^\nu\Psi,\label{main:eq5:Z}\\
&& \nabla_\mu C_\nu-\nabla_\nu C_\mu-[C_\mu,C_\nu] = C_{\mu\nu},\label{main:eq2:Z}\\
&& \nabla_\mu C^{\mu\nu}-[C_\mu,C^{\mu\nu}]=\pi_-(i K^\nu),\label{main:eq3:Z}\\
&&\nabla_\mu i K^\nu - [C_\mu, i K^\nu]= i\alpha(4\,\tau_\mu\tau^\nu-\delta^\nu_\mu)\Psi\Psi^\dagger
\label{main:eq6:Z}
\end{eqnarray}
в которой неизвестными являются $\Psi, K^\mu, C_\mu, A_\mu$ (вещественное скалярное поле $\alpha=\alpha(x)$ предполагается известным, но  интерпретацию этого скалярного  поля пока оставляем в стороне).

Отметим, что из уравнения (\ref{main:eq6:Z}) с помощью формулы
(\ref{CmuViaKnu}) можно выразить компоненты ковектора $C_\mu$ через $K^\nu$, $\Psi$ и подставить результат в уравнения (\ref{main:eq2:Z}), (\ref{main:eq3:Z}). В результате получим одно уравнение, связывающее $K^\nu$ и $\Psi$. То есть получается связь между метрическим тензором (квадратным корнем $K^\nu$ из метрического тензора) и волновой функцией фермиона $\Psi$.
 Ввиду громоздкости,  это уравнение не выписываем.

%Условно можно считать, что уравнение (\ref{main:eq1:Z}) является уравнением для определения $\Psi$; уравнения (\ref{main:eq4:Z}), (\ref{main:eq5:Z}) являются уравнениями для определения электрослабого поля $(A_\mu, A_{\mu\nu})$; уравнения (\ref{main:eq2:Z}), (\ref{main:eq3:Z}), (\ref{main:eq6:Z}) являются уравнениями для определения гравитационного поля $(C_\mu, C_{\mu\nu})$ и корня из метрического тензора $K^\mu$.

Из уравнений (\ref{main:eq2:Z}),(\ref{main:eq3:Z}) в качестве следствия вытекает равенство
$$
\nabla_\mu\pi_-(i K^\mu) - [C_\mu, \pi_-(i K^\mu)]=0,
$$
которое вместе с равенствами (\ref{cons:kpsi:Z}) и $\nabla_\mu\pi_+( i K^\mu)=0$ дает равенство
$$
\nabla_\mu(\Psi^\dagger i K^\mu\Psi) - [A_\mu, \Psi^\dagger i K^\mu\Psi]=0.
$$

Система уравнений (\ref{main:eq1:Z})-(\ref{main:eq6:Z}) является инвариантной относительно калибровочных преобразований (\ref{gauge:Lanczos1:Z}), (\ref{gauge:Lanczos2:Z}) и (\ref{gauge:Lanczos3:Z}). При этом для преобразования (I) имеем
$$
C_{\mu\nu}\to\acute C_{\mu\nu}=C_{\mu\nu},\quad A_{\mu\nu}\to\acute A_{\mu\nu}=V^{-1}A_{\mu\nu}V,
$$
для преобразования (II) имеем
$$
C_{\mu\nu}\to\acute C_{\mu\nu}=S^{-1}C_{\mu\nu} S,\quad A_{\mu\nu}\to\acute A_{\mu\nu}=S^{-1}A_{\mu\nu}S,
$$
и  для преобразования (III) имеем
$$
C_{\mu\nu}\to\acute C_{\mu\nu}=S^{-1}C_{\mu\nu} S,\quad A_{\mu\nu}\to\acute A_{\mu\nu}=A_{\mu\nu}.
$$

Наконец отметим возможность рассматривать систему уравнений
(\ref{main:eq1:Z}) -- (\ref{main:eq6:Z}) из которой выброшено уравнение (\ref{main:eq6:Z}), либо выброшены уравнения (\ref{main:eq2:Z}), (\ref{main:eq3:Z}). Анализ этих вариантов представляется полезным.

%%%%%%%%%%%%%%%%%%%%%%%%%%%%%%%%%%%%%%%%%%%%%%%%%%%%%%%%%%%%%%%%%%%%%%%%%%

\section{Основная система уравнений с $\SU(2)$, $\U(2)$, $\U(3)$ калибровочной симметрией}

Выпишем систему уравнений с тремя неабелевыми калибровочными симметриями с группами Ли $\SU(2)$, $\U(2)$, $\U(3)$.  Эту систему уравнений будем рассматривать как заготовку для теории, которая, по нашему мнению, должна воспроизводить Стандартную Модель электрослабых и сильных (цветовых, $\SU(3)$ калибровочная симметрия) взаимодействий кварков с учетом гравитационного взаимодействия кварков ($\SU(2)$ калибровочная симметрия). Отметим, что для этого нам не потребуется вводить $\SU(5)$ калибровочную симметрию.
\medskip

Введем несколько математических структур.

1) Скалярные и тензорые поля на многообразии $\M$ со значениями в $\MC$: вектор $K^\mu\in\Herm(2)\T^1$, удовлетворяющий условиям (\ref{cond:kmu1}), (\ref{cond:kmu2});
$\Psi_1,\Psi_2,\Psi_3\in\MC$; $A_\mu\in \u(2)\T_1$, $A_{\mu\nu}\in \u(2)\T_2$; $C_\mu\in \su(2)\T_1$, $C_{\mu\nu}\in \su(2)\T_2$;

2) $3\times3$-матрицы из $\Mat(3,\C)$ будем обозначать подчеркиванием: $\underline{T},\underline{E},\underline{B_\mu}\in\Mat(3,\C)$, где
$$
\underline{T}=\begin{pmatrix}1&0&0\cr 0&0&0\cr 0&0&0\end{pmatrix},
\quad
\underline{E}=\begin{pmatrix}1&0&0\cr 0&1&0\cr 0&0&1\end{pmatrix},
\quad
\underline{B_\mu}=\begin{pmatrix}b^1_{1\mu}&b^1_{2\mu}&b^1_{3\mu}\cr
b^2_{1\mu}&b^2_{2\mu}&b^2_{3\mu}\cr b^3_{1\mu}&b^3_{2\mu}&b^3_{3\mu}
\end{pmatrix},
$$

3) Рассматриваем тензорное произведение  $\Mat(3,\C)\otimes\MC$.
Элементы тензорного произведения  $\Mat(3,\C)\otimes\MC$ обозначаем знаком $\check{}$ и рассматриваем как матрицы со значениями в алгебре $\MC$, либо как элементы алгебры $\MC$ со значениями  в матрицах (из $\Mat(3,\C)$)
$$
\check \Psi=(\underline{T}\otimes e)\check\Psi=\begin{pmatrix}\Psi_1&\Psi_2&\Psi_3\cr 0&0&0\cr 0&0&0\end{pmatrix},
$$
$$
\check A_\mu=\underline{E}\otimes A_\mu=\begin{pmatrix}A_\mu&0&0\cr 0&A_\mu&0\cr 0&0&A_\mu\end{pmatrix},
$$
$$
\check B_\mu=\underline{B_\mu}\otimes e=\begin{pmatrix}b^1_{1\mu}e&b^1_{2\mu}e&b^1_{3\mu}e\cr
b^2_{1\mu}e&b^2_{2\mu}e&b^2_{3\mu}e\cr b^3_{1\mu}e&b^3_{2\mu}e&b^3_{3\mu}e
\end{pmatrix},
$$
$$
\check K^\mu=\underline{E}\otimes K^\mu,\quad
\check C_\mu=\underline{E}\otimes C_\mu.
$$

Теперь выпишем модельное уравнение типа Дирак-Ланцоша
\begin{equation}
i\check K^\mu(\nabla_\mu\check \Psi + \check \Psi \check A_\mu + \check \Psi \check B_\mu - \check C_\mu\check \Psi)=0.\label{U223:eq}
\end{equation}
Это уравнение можно переписать без матриц третьего порядка в виде системы трех уравнений в алгебре матриц второго порядка
\begin{equation}
i K^\mu(\nabla_\mu\Psi_l +\Psi_l A_\mu +\Psi_k b^k_{l\mu} -  C_\mu \Psi_l)=0,\quad l=1,2,3\label{U223:eq1}
\end{equation}
где по индексу $k$ ведется суммирование от $1$ до $3$.

Выпишем калибровочные симметрии уравнения (\ref{U223:eq}).
%Алгебра Ли $L(t)=\{\theta t, e^{12}t, e^{13}t, e^{23}t\}$.

1) Умножение справа на $\check U=\underline{E}\otimes U$, $U\in \U(2)$
\begin{eqnarray*}
\check\Psi &\to& \check\Psi\check U,\\
\check K^\mu &\to& \check K^\mu,\\
\check A_\mu &\to& \check{U}^{-1}\check A_\mu\check U-  \check{U}^{-1}\nabla_\mu\check{U}=
\underline{E}\otimes(U^{-1}A_\mu U-U^{-1}\nabla_\mu U),\\
\check B_\mu &\to& \check B_\mu,\\
\check C_\mu &\to& \check C_\mu;
\end{eqnarray*}

2) Умножение справа на $\check V=\underline{V}\otimes e$, $\underline{V}\in \U(3)$
\begin{eqnarray*}
\check\Psi &\to& \check\Psi\check V,\\
\check K^\mu &\to& \check K^\mu,\\
\check A_\mu &\to& \check A_\mu,\\
\check B_\mu &\to& \check{V}^{-1}\check B_\mu\check V-  \check{V}^{-1}\nabla_\mu\check{V}=
(\underline{V}^{-1}\underline{B}_\mu \underline{V}-\underline{V}^{-1}\nabla_\mu \underline{V})\otimes e,\\
\check C_\mu &\to& \check C_\mu;
\end{eqnarray*}

3) Умножение слева на $\check{S}^{-1}=\underline{E}\otimes S^{-1}$, $S\in \SU(2)$ и справа на $\check{S}=\underline{E}\otimes S$
\begin{eqnarray*}
\check\Psi &\to& \check{S}^{-1}\check\Psi\check S,\\
\check K^\mu &\to& \check{S}^{-1}\check K^\mu\check{S}=\underline{E}\otimes(S^{-1}K^\mu S),\\
\check A_\mu &\to& \check{S}^{-1}\check A_\mu\check S- \check{S}^{-1}\nabla_\mu\check{S}=
\underline{E}\otimes(S^{-1}A_\mu S- S^{-1}\nabla_\mu S),\\
\check B_\mu &\to& \check B_\mu,\quad\mbox{($[\check S, \check{B}_\mu]=0$)},\\
\check C_\mu &\to& \check{S}^{-1}\check C_\mu\check S- \check{S}^{-1}\nabla_\mu\check{S}=
\underline{E}\otimes(S^{-1}C_\mu S- S^{-1}\nabla_\mu S).
\end{eqnarray*}
Определим вектор $\check J^\mu$ со значениями в тензорном произведении $\Mat(3,\C)\otimes\MC$
$$
\check J^\mu := \check \Psi^\dagger i\check K^\mu\check \Psi = \|J^\mu_{kl}\|,
$$
где
$$
J^\mu_{kl} = \Psi^\dagger_k i K^\mu\Psi_l.
$$
Обозначим через $\check\Psi^\dagger$ эрмитово сопряженный элемент тензорного произведения
$$
\check \Psi^\dagger=\check\Psi^\dagger(\underline{T}\otimes e)=\begin{pmatrix}\Psi_1^\dagger&0&0\cr\Psi_2^\dagger&0&0\cr \Psi_3^\dagger&0&0\end{pmatrix},
$$
Уравнение (\ref{U223:eq}) слева умножим на $\check\Psi^\dagger$ и вычтем эрмитово сопряженное выражение. Получим
\begin{equation}
(\nabla_\mu \check J^\mu - [\check A_\mu+\check B_\mu,\check J^\mu])-
\check \Psi^\dagger(\nabla_\mu(i\check K^\mu)-[\check C_\mu,i\check K^\mu])\check\Psi=0.
 \label{1st:cons}
\end{equation}
Возьмем  операцию взятия (матричного) следа $\Tr\,:\,\Mat(3,\C)\otimes\MC\to\MC$ и определим
$$
J^\mu_{(A)} :=\frac{1}{3}\Tr\,\check J^\mu = \frac{1}{3}(J^\mu_{11}+J^\mu_{22}+J^\mu_{33}).
$$
Отметим, что $\Tr([\check B_\mu,\check J^\mu])=0$. Соответственно, взяв от обеих частей равенства $(\ref{1st:cons})$ операцию $\frac{1}{3}\Tr$, получим следствие
\begin{equation}
(\nabla_\mu J_{(A)}^\mu - [A_\mu,J_{(A)}^\mu]) - \frac{1}{3}\Tr\big(\check \Psi^\dagger(\nabla_\mu(i\check K^\mu)-[\check C_\mu,i\check K^\mu])\check\Psi
\big)=0.\label{2nd:cons}
\end{equation}
Возьмем операцию проектирования на подпространство элементов, натянутых на единичную матрицу второго порядка $e$
$$
\pi^0\,:\,\Mat(3,\C)\otimes\MC\to\C\times e
$$
и операцию
$$
\dot\pi^0\,:\,\Mat(3,\C)\otimes\MC\to\C,\quad \dot\pi^0(U)=\pi^0(U)|_{e\to1}.
$$

Тогда
$$
\underline{J}^\mu_{(B)}:=\dot\pi^0(\check J^\mu) = \begin{pmatrix}
\dot\pi^0(J^\mu_{11}) & \dot\pi^0(J^\mu_{12}) & \dot\pi^0(J^\mu_{13})\cr
\dot\pi^0(J^\mu_{21}) & \dot\pi^0(J^\mu_{22}) & \dot\pi^0(J^\mu_{23})\cr
\dot\pi^0(J^\mu_{31}) & \dot\pi^0(J^\mu_{32}) & \dot\pi^0(J^\mu_{33})
\end{pmatrix}\in\Mat(3,\C).
$$
Не трудно убедиться, что
$$
\dot\pi^0([\check B_\mu,\check J^\mu]) = [\underline B_\mu,\dot\pi^0(\check J^\mu)].
$$
Поэтому, подействовав на обе части равенства (\ref{1st:cons}) оператором $\dot\pi^0$, получим
\begin{equation}
(\nabla_\mu\underline{J}_{(B)}^\mu - [B_\mu,\underline{J}_{(B)}^\mu]) - \pi^0\big(\check \Psi^\dagger(\nabla_\mu(i\check K^\mu)-[\check C_\mu,i\check K^\mu])\check\Psi
\big)=0.\label{3d:cons}
\end{equation}

Таким образом, доказана следующая теорема.
\begin{theorem}\label{theorem5}
Если выполнены условия
\begin{eqnarray}
&& \nabla_\mu J^\mu_{(A)} - [A_\mu, J^\mu_{(A)}]=0,\label{J:A}\\
&& \nabla_\mu\underline{J}^\mu_{(B)} - [\underline{B}_\mu, \underline{J}^\mu_{(B)}]=0,\label{J:B}\\
&& \nabla_\mu J^\mu_{(C)} - [C_\mu, J^\mu_{(C)}]=0,\label{J:C}
\end{eqnarray}
где
$$
J^\mu_{(A)} = \frac{1}{3}(J^\mu_{11} +J^\mu_{22} +J^\mu_{33}),\quad
\underline{J}^\mu_{(B)}=\dot\pi^0(\check J^\mu),\quad
J^\mu_{(C)} = \pi_-(i K^\mu),
$$
и
$$
\check J^\mu= \begin{pmatrix}
J^\mu_{11} & J^\mu_{12} & J^\mu_{13}\cr
J^\mu_{21} & J^\mu_{22} & J^\mu_{23}\cr
J^\mu_{31} & J^\mu_{32}&  J^\mu_{33}
\end{pmatrix}=\check\Psi^\dagger i\check K^\mu\check\Psi,
$$
то выполняются равенства (\ref{2nd:cons}) и (\ref{3d:cons}).
\end{theorem}
\medskip

Постулируем уравнения для $A_\mu,\underline{B}_\mu,C_\mu$
\begin{eqnarray}
\nabla_\mu A_\nu - \nabla_\nu A_\mu - [A_\mu, A_\nu] = A_{\mu\nu},\nonumber\\
\nabla_\mu A^{\mu\nu} - [A_\mu, A^{\mu\nu}] = J^\nu_{(A)},\label{YM:A}\\
\nabla_\mu\underline{B}_\nu - \nabla_\nu\underline{B}_\mu - [\underline{B}_\mu, \underline{B}_\nu] = \underline{B}_{\mu\nu},\nonumber\\
\nabla_\mu \underline{B}^{\mu\nu} - [\underline{B}_\mu, \underline{B}^{\mu\nu}] = \underline{J}^\nu_{(B)},\label{YM:B}\\
\nabla_\mu C_\nu - \nabla_\nu C_\mu - [C_\mu, C_\nu] = C_{\mu\nu},\nonumber\\
\nabla_\mu C^{\mu\nu} - [C_\mu, C^{\mu\nu}] = J^\nu_{(C)},\label{YM:C}
\end{eqnarray}
Это есть три пары уравнений Янга-Миллса для потенциалов $A_\mu,\underline{B}_\mu,C_\mu$ и напряженностей $A_{\mu\nu},\underline{B}_{\mu\nu},C_{\mu\nu}$ с неабелевыми токами $J^\nu_{(A)}, \underline{J}^\nu_{(B)}, J^\nu_{(C)}$.

Из этих уравнений Янга-Миллса вытекают следствия (\ref{J:A}), (\ref{J:B}), (\ref{J:C}). Поэтому уравнения Янга-Миллса (\ref{YM:A}), (\ref{YM:B}), (\ref{YM:C}) согласованы с модельным уравнением (\ref{U223:eq}) в том смысле, что из выполнения равенств (\ref{J:A}), (\ref{J:B}), (\ref{J:C}) следует выполнение равенства (\ref{1st:cons}) (сравни с Теоремой 1).

Замечание. Обратим внимание читателя на возможность таких модификаций рассматриваемой модели, при которых для определения потенциала  $C_\mu$ в дополнение к уравнениям Янга-Миллса (\ref{YM:C}) или вместо уравнений Янга-Миллса (\ref{YM:C}) используется {\em примитивное полевое уравнение}
\begin{equation}
\nabla_\mu J^\nu_{(C)} - [C_\mu, J^\nu_{(C)}]=\alpha(\tau^\nu\tau_\mu-
\delta^\nu_\mu)\,\Tr(\check\Psi{\check\Psi}^\dagger),
\quad \mu,\nu=0,1,2,3.\label{Jmunu:C}
\end{equation}
Это уравнение является инвариантным относительно всех трех указанных калибровочных преобразований 1), 2), 3) и, в частности, относительно калибровочного преобразования
$$
J^\nu_{(C)} \to S^{-1} J^\nu_{(C)} S,\quad C_\mu\to S^{-1}C_\mu S-S^{-1}\nabla_\mu S,
$$
$$
\Psi_k\to S^{-1}\Psi_k S,\quad k=1,2,3,
$$
где $S=S(x)\in \SU(2)$. При этом $C_\mu\in \su(2)\T_1$, $J^\nu_{(C)}\in \su(2)\T^1$, $\Psi_k\in\Mat(2,\C)$.

%%%%%%%%%%%%%%%%%%%%%%%%%%%%%%%%%%%%%%%%%%%%%%%%%%%%%

\section{Сравнение результатов настоящей статьи с результатами предшествующей статьи}\label{par:AACA}

В настоящей статье мы развиваем и модифицируем результаты статьи \cite{Marchuk_AACA2021}. В этом параграфе мы покажем как свести  уравнение (12) из \cite{Marchuk_AACA2021} к уравнению дираковского типа (\ref{mod:Lanczos:Z}), которое в этом параграфе рассматриваем в пространстве Минковского $\R^{1,3}$ ($g_{\mu\nu}=\eta_{\mu\nu}$ и $\nabla_\mu=\partial_\mu$).

В статье \cite{Marchuk_AACA2021} рассматривались тензорные поля в пространстве Минковского $\R^{1,3}$  со значениями в комплексифицированной алгебре Клиффорда $\Ccl$ с единицей $e$, с порождающими $e^a$, $a=0,1,2,3$ такими, что $e^a e^b+e^b e^a=2\eta^{ab}e$.

Уравнение (12) из \cite{Marchuk_AACA2021} имеет вид\footnote{В этом уравнении мы ставим штрихи над $\Psi,A_\mu,C_\mu$ чтобы отличать их от соответствующих величин уравнения (\ref{mod:Lanczos:Z}).}
\begin{equation}
h^\mu(\partial_\mu\acute\Psi- \acute C_\mu\acute\Psi + \acute\Psi\acute A_\mu) + i m\acute\Psi=0,
\label{eq:AACA2031}
\end{equation}
где $h^\mu$ -- генвектор (то есть $h^\mu h^\nu+h^\nu h^\mu=2\eta^{\mu\nu}e$ и $h^\mu=y^\mu_a e^a$); $\acute\Psi=\acute\Psi\chi\in I(\chi)\subset\Ccl$; $\chi=(e-i\theta)/2$; $\acute A_\mu\in L(\chi)\T_1$; $\acute C_\mu\in L_3\T_1$; $\beta=e^0$, $\theta=e^0e^1e^2e^3$, $\tau^1=e^2 e^3$, $\tau^2=-e^1e^3$, $\tau^3=e^1e^2$ и
\begin{eqnarray*}
I(\chi) &=& \{U\in\Ccl : U=U\chi\},\\
N(\chi) &=& \{U\in I(\chi) : U=\chi U\},\\
L(\chi) &=& \{S\in N(\chi) : S^\dagger=-S\},\\
L_3 &=& \{S\in\cl^\Even : S^\dagger=-S,\ [\beta,S]=0\},\\
L_4 &=& \{S\in\cl^\Even : S^\dagger=-S\}.
\end{eqnarray*}
Легко видеть, что $L(\chi)$ -- вещественное четырехмерное векторное пространство, натянутое на базисные элементы $\tau^1\chi$, $\tau^2\chi$, $\tau^3\chi$, $\theta\chi$. Если $S\in L_4$, то $R=S\chi\in L(\chi)$.

Уравнение (\ref{eq:AACA2031}) умножим слева на $i\chi\beta$. Получим при $m=0$
$$
i\beta h^\mu\chi(\partial_\mu(\chi\acute\Psi) - \acute C_\mu(\chi\acute\Psi) + (\chi\acute\Psi)A_\mu))=0.
$$
Так как $[\acute C_\mu,\chi]=0$ и $\chi^2=\chi$, то $\acute C_\mu(\chi\acute\Psi)=
(\chi\acute C_\mu)(\chi\acute\Psi)$ и можно ввести величины $\check C_\mu=\chi\acute C_\mu\in N(\chi)\T_1$, $\check\Psi=\chi\acute\Psi\in N(\chi)$, $\check K^\mu=\beta h^\mu\chi=\chi\beta h^\mu\chi\in N(\chi)\T^1$,
$\check A_\mu=\acute A_\mu\in N(\chi)\T_1$, которые удовлетворяют уравнению
\begin{equation}
i\check K^\mu(\partial_\mu\check\Psi - \check C_\mu\check\Psi + \check\Psi\check A_\mu)=0\label{chK:eq}
\end{equation}
и все величины входящие в это уравнение принадлежат $N(\chi)$ -- пересечению правого и левого идеалов \cite{Riesz}, порожденных идемпотентом $\chi$.

Если предположить, что исходное уравнение (\ref{eq:AACA2031}) описывает волновую функцию лептона, то, положив $m=0$ и перейдя в правый идеал (умножением слева на $\chi$), мы получим уравнение для волновой функции, которую можно интерпретировать как волновую функцию левоспиральной частицы (лептона) в соответствии с представлениями Стандартной Модели.

Наконец, запишем уравнение (\ref{chK:eq}) как матричное уравнение. С этой целью возьмем для порождающих $e^a$ алгебры Клиффорда $\Ccl$ такое представление в виде матриц $T : \Ccl\to \Mat(4,\C)$, что
$$
T(e)=\diag(1,1,1,1),\quad T(e^{0123})=\diag(i,i,-i,-i),\quad T(\chi)=\diag(1,1,0,0).
$$
В этом случае множество матриц $T(N(\chi))$ есть множество блочно-диагональных комплексных матриц вида
$$
\begin{pmatrix}
u_{11} & u_{12} & 0 & 0\cr
u_{21} & u_{22} & 0 & 0\cr
0 & 0 & 0 & 0 \cr
0 & 0 & 0 & 0
\end{pmatrix}.
$$
Поэтому имеет место изоморфизм алгебр
$$
T(N(\chi))\simeq \Mat(2,\C).
$$
Переходя в уравнении (\ref{chK:eq})
к описанному матричному представлению, придем к уравнению (\ref{mod:Lanczos:Z}). Таким образом мы показали как уравнение (12) из \cite{Marchuk_AACA2021} свести к модельному уравнению Дирака-Ланцоша (\ref{mod:Lanczos:Z}).

%%%%%%%%%%%%%%%%%%%%%%%%%%%%%%%%%%%%%%%%%%%%%%%%%%%%%%
\section{Заключительные замечания}

В статье дан эскиз модели, объединяющей гравитационные, электрослабые и сильные (КХД) взаимодействия элементарных частиц. Пока лишь грубая схема -- модельное уравнение Дирака-Ланцоша для частиц спина $1/2$; уравнения Янга-Миллса  для частиц спина $1$, рассматриваемых как медиаторы электрослабых, сильных и гравитационных взаимодействий; плюс примитивное полевое уравнение, связывающее гравитационное поле с метрикой многообразия (корнем из метрического тензора).

 Контуры модели уже видны, но детализацию еще предстоит разработать. Если говорить о детализации той части модели, которая должна описывать электрослабые и сильные взаимодействия частиц, то в целом понятно, что воспроизводить надо хорошо обкатанную схему Стандартной Модели. Детализация гравитационной части модели не столь очевидна, так как  требует, в конечном счете, соотнесения с ОТО А.~Эйнштейна.

В этой связи обсудим вопрос, а почему вообще считаем, что дополнительная $\SU(2)$ калибровочная симметрия, появившаяся в модельном уравнении Дирака-Ланцоша, описывает гравитацию. Один аргумент состоит в том, что указанная $\SU(2)$ калибровочная симметрия действует на все поля, присутствующие в модели (см. формулы (\ref{gauge:Lanczos2:Z})), что мы и ожидаем от гравитации. Других аргументов у меня пока нет. Эту ситуацию можно выразить такими словами: в статье обсуждается гипотеза о том, что гравитация описывается (янг-миллсовским) калибровочным полем с $\SU(2)$  симметрией.

 Планируется дальнейшее изучение этих вопросов, а также вопросов, связанных с лагранжевой и гамильтоновой формулировкой модели, квантованием и др.

%В рассматриваемой модели используется ряд результатов и идей, сформировавшихся усилиями многих исследователей в релятивистской квантовой физике (в частности, в теории калибровочных полей) и в теории гравитации.  Представляется, что модель содержит оригинальное сочетание известных и новых идей (таких как модельное уравнение Дирака-Ланцоша с двумя унитарными неабелевыми калибровочными симметриями, примитивное полевое уравнение и др.) и это дает надежду на то, что при дальнейшей разработке модель разовьется до полноценной теории поля.

%Думается, что такая перспектива оправдывает время и усилия, затраченные автором на создание модели.

%Вместе с тем, нельзя исключать того, что реалистичная единая теория поля будет создана на основе других идей, разрабатываемых другими исследователями. В любом случае надеюсь, что наличие еще одной модели будет вкладом в копилку новых идей из которой, в конце концов, выкристаллизуется правильная единая теория поля.

Приведем список новых математических результатов, содержащихся в настоящей статье.
\begin{enumerate}
\item Введено модельное уравнение Дирака-Ланцоша (\ref{mod:Lanczos:Z}) с двумя неабелевыми унитарными калибровочными симметриями относительно групп Ли $\SU(2)$ и $\U(2)$ (Теорема \ref{theorem4}) и система уравнений (\ref{U223:eq}), (\ref{U223:eq1}) с тремя неабелевыми унитарными калибровочными симметриями относительно групп Ли $\SU(2)$, $\U(2)$ и $\U(3)$.

\item Введено примитивное полевое уравнение с общей правой частью (\ref{prim:eq:mat}) и примитивное полевое уравнение (\ref{primitive:fe}) с правой частью, зависящей от волновой функции фермиона $\Psi$. Доказана калибровочная симметрия относительно групп Ли $\SU(2)$ и $\U(2)$, а также калибровочная симметрия уравнения (\ref{Jmunu:C}) относительно групп Ли $\SU(2)$, $\U(2)$ и $\U(3)$.  Найдено решение примитивного уравнения (\ref{CmuViaKnu}) (Теорема \ref{theorem2}).

\item Введена система уравнений (\ref{main:eq1:Z}) -- (\ref{main:eq6:Z}) которую, после необходимой детализации (угол Вайнберга, левоспиральность, механизм Хиггса и др.), можно рассматривать как систему уравнений, описывающую лептоны (волновая функция $\Psi$), взаимодействующие одновременно с электрослабым полем (калибровочная симметрия $\U(2)$) и с гравитационным полем (калибровочная симметрия $\SU(2)$).
    %Отметим, что все физические константы пока взяты равными единице.

\item Введена система уравнений (\ref{U223:eq}), (\ref{YM:A}), (\ref{YM:B}), (\ref{YM:C}), (\ref{Jmunu:C}) которую, после необходимой детализации,  можно рассматривать как систему уравнений, описывающую кварки, взаимодействующие одновременно с электрослабым полем (калибровочная симметрия $\U(2)$), с цветовым полем КХД (калибровочная симметрия $\U(3)$)
     и с гравитационным полем (калибровочная симметрия $\SU(2)$).
 \end{enumerate}

%%%%%%%%%%%%%%%%%%%%%%%%%%%%%%%%%%%%%%%%%%%%%%%%%%%%%%
%\end{fulltext}

\medskip

Марчук Николай Гурьевич,
\medskip

Адрес и аффилиация:

1) Москва, 119991, ул. Губкина д.8,

 Математический институт им. В.А.Стеклова РАН, отдел математической физики

2) Москва, 109028, Покровский бульвар, д.11,

Национальный исследовательский университет ``Высшая школа экономики'',
Факультет экономических наук, Департамент математики
\medskip

phone: +7-916-1845025

\medskip

email: nmarchuk@mi-ras.ru

\medskip

Marchuk Nikolay

Address and affiliation:

1) Moscow, 119991, st. Gubkina 8,

  Steklov Mathematical Institute of Russian Academy of Sciences, department of mathematical physics

2) Moscow, 109028, Pokrovsky Boulevard, 11,

National Research University ``Higher School of Economics'',
Faculty of Economic Sciences, Department of Mathematics
\medskip

Title:

Sketch of a gauge model of gravity with SU(2) symmetry on a Lorentzian manifold with tetrad
\medskip

Abstract:

A gauge model with SU(2) symmetry is proposed to describe the gravitational interaction of fundamental fermions (leptons and quarks) on a Lorentzian manifold with a tetrad.  From the system of Dirac-Yang-Mills equations underlying the Standard Model, we arrive at a model system of Dirac-Lanczos-Yang-Mills equations, written using second-order matrices. This system of equations has an additional gauge symmetry with respect to the unitary group SU(2). The Yang-Mills field associated with this gauge group is interpreted as the gravitational field interacting with fundamental fermions.
\medskip

Key words:

General Relativity, gravity, Lorentzian manifold, tetrad, Pauli matrices, Dirac equation, Yang-Mills equations, Dirac-Lanczos equation, gauge symmetry, primitive field equation

\end{document}